\documentclass[aps,prl,twocolumn,superscriptaddress,longbibliography,nobalancelastpage,floatfix]{revtex4-2}
\usepackage[T1]{fontenc}
\usepackage[utf8]{inputenc}
\pdfoutput=1 
\usepackage{graphicx}
\usepackage{amsmath}
\usepackage{amssymb}
\usepackage[explicit]{titlesec} 
\usepackage{bm}
\usepackage{physics}
\usepackage[colorlinks = true,
            allcolors = {Blue}]{hyperref}
\usepackage[capitalize]{cleveref}
\usepackage[caption=false]{subfig}
\newcommand{\phantomsubfloat}[1]{
    {%
        \captionsetup[subfigure]{labelformat=empty}
        \subfloat[][]{#1}
    }%
}
\usepackage[dvipsnames]{xcolor}

\newcommand{\new}[1]{#1}
\newcommand{\rhospec}{\varrho}
\newcommand{\vect}[1]{\bm{#1}}

\begin{document}

\author{Piper Fowler-Wright}
\affiliation{SUPA, School of Physics and Astronomy, University of St Andrews, St Andrews, KY16 9SS, United Kingdom}
\author{Brendon W. Lovett}
\affiliation{SUPA, School of Physics and Astronomy, University of St Andrews, St Andrews, KY16 9SS, United Kingdom}
\author{Jonathan Keeling}
\affiliation{SUPA, School of Physics and Astronomy, University of St Andrews, St Andrews, KY16 9SS, United Kingdom}

\title{Efficient Many-Body Non-Markovian Dynamics of Organic Polaritons} 
\date{\today} 
\begin{abstract} 

	We show how to simulate a model of many molecules with both strong coupling
	to many vibrational modes and collective coupling to a single photon mode.
	We do this by combining process tensor matrix product operator methods with
	a mean-field approximation which reduces the dimension of the problem.  We
	analyze the steady state of the model under incoherent pumping to determine
	the dependence of the polariton lasing threshold on cavity detuning,
	light-matter coupling strength, and environmental temperature. 
	Moreover, by measuring two-time correlations, we study quadratic fluctuations
	about the mean field to calculate the photoluminescence spectrum.
	Our method enables one to simulate many-body systems with strong coupling to
	multiple environments, and to extract both static and dynamical properties.  
	
	\vspace{0.8\baselineskip}
	\noindent
	\small{DOI: \href{https://doi.org/10.1103/PhysRevLett.129.173001}{10.1103/PhysRevLett.129.173001}}
\end{abstract}

\maketitle

The strong coupling between organic matter confined in a microcavity and light
results in new collective modes---superpositions of molecular excitations and
photons known as exciton polaritons~\cite{Keeling2020}. Under sufficient
pumping, these may condense into a coherent or lasing state, as has now been
demonstrated in a diverse range of organic materials~\cite{Kena2010,
Plumhof2014, Daskalakis2014, Grant2016, Dietrich2016,Cookson2017} (see
Ref.~\cite{Keeling2020} for a review).  The rich photophysics of organic
molecules allows for the possibility of room temperature lasing devices with
ultralow thresholds, yet also makes the task of determining the optimal
conditions for lasing a challenging one. In particular, one must consider the
effect on the dynamics of the vibrational environment of each
molecule~\cite{Herrera2018}, which is generally structured and beyond weak
coupling or Markovian treatments~\cite{Thorwart2009, Ishizaki2009,Prior2010,
Fassioli2012,Chin2013,IlesSmith2016, DelPino2018, DelPino2018a, Clear2020}.  To
this end there have been studies of polariton condensation using simplified
models with a few vibrational modes~\cite{Cwik2014, Spano2015, Galego2015,
Herrera2016, Wu2016, Herrera2016a, Herrera2016b, Herrera2018, Zeb2018,
Strashko2018, Arnardottir2020}, and also studies involving exact vibrational
spectra for a small number of molecules~\cite{DelPino2018,DelPino2018a}.
However, the real system has both a complex vibrational density of states and
many, e.g., \(10^5\), molecules. Therefore, what is needed is a method capable of
handling large systems with non-Markovian effects.  Here we provide such
a method and show the consequences for the description of polariton lasing.

Process tensor matrix product operator (PT-MPO) methods are a class of numerical
methods based upon the process tensor (PT) description of open quantum system
dynamics~\cite{Strathearn2018,Pollock2018, Jorgensen2019, Fux2021,
Cygorek2022,Bose2021, Richter2022}.  The PT captures all
possible effects of the environment on a system. 
The system Hamiltonian
propagator, or any system operator, then forms a finite set of interventions
that may be contracted with the PT and thus one can find any system
observable or multitime correlation function.  Crucially, the PT can be
represented efficiently as a matrix product operator that only needs to be
calculated once for a given system-bath interaction and set of bath
conditions~\cite{Fux2021}.  While this provides an efficient
means to evolve a system with long memory times, such methods have so far been
limited to systems of small Hilbert space dimension.

\begin{figure}
	\centering
    \vspace{-2\baselineskip}%
	\phantomsubfloat{\label{fig:1a}}
    \phantomsubfloat{\label{fig:1b}}
    \phantomsubfloat{\label{fig:1c}}
    \phantomsubfloat{\label{fig:1d}}

	\includegraphics[scale=.395]{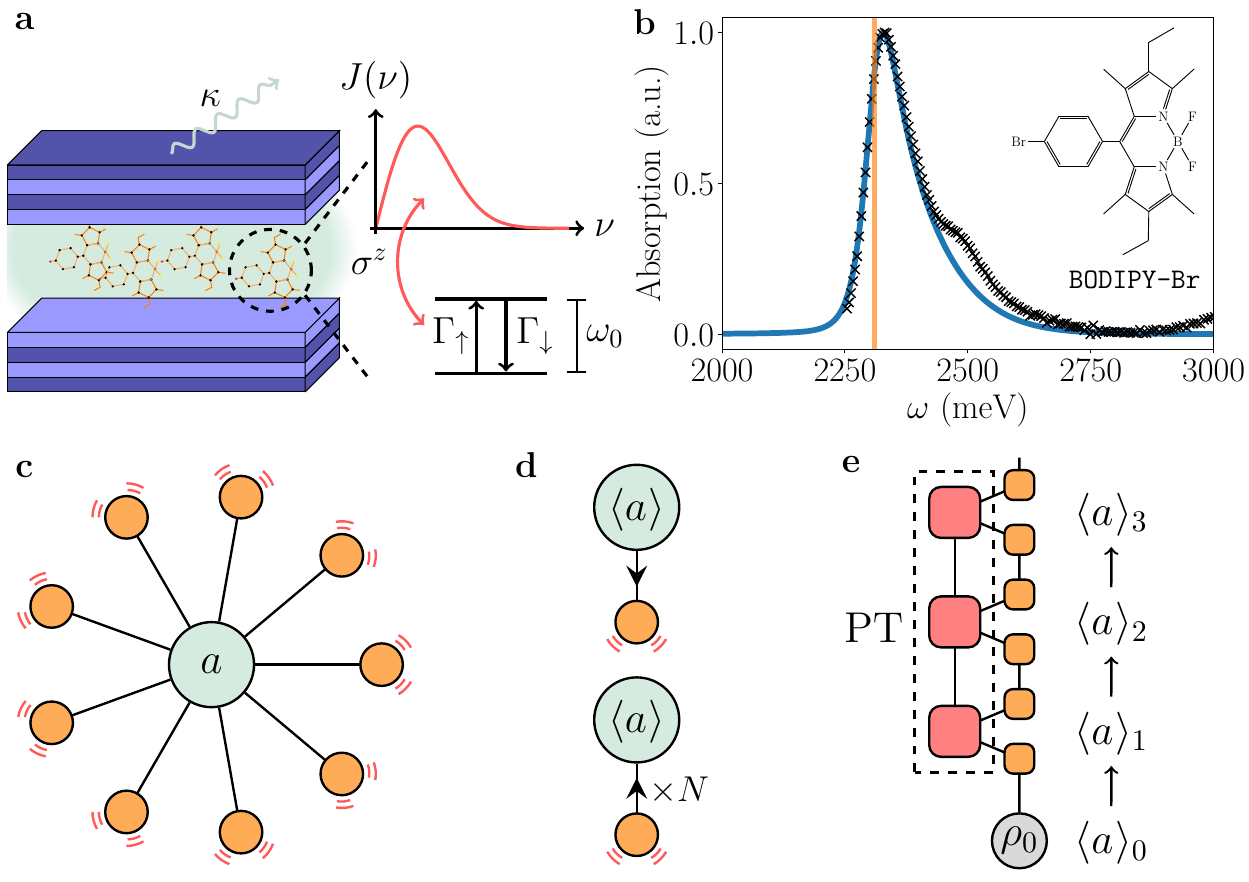}
	\caption{(a) 
	Our system: a molecular ensemble in an optical microcavity. 
		Each molecule is modeled as a driven-dissipative two-level
		system with a diagonal coupling to a harmonic environment.  
		The spectral density \(J(\nu)\) of the environment is chosen to match (b)
		absorption data~\cite{Grant2016} for BODIPY-Br at
		\(300\)K (black crosses: experimental data, blue curve: model spectrum,
		orange line: \(\omega_0=2310\)~meV).  For the Ohmic form \cref{eq:J}
		with dissipation \(\Gamma_\downarrow = 10\)~meV we obtained
		\(\alpha=0.25\) and \(\nu_c=150\)~meV (\(\hbar=1\)). 
		(c) Schematic of the many-body open
		system before and (d) after the mean-field reduction.
		(e) Tensor network for the PT-MPO method with the concurrent dynamics of
		the cavity field.  The PT (red)	is constructed independently of the
		system propagators	(orange) and initial state (gray), allowing the
		dynamics for many different system Hamiltonians to be calculated at
		relatively little cost.}
\label{fig:1} \end{figure}

In this Letter we present a mean-field approach to reduce an \(N\)-body problem
to one that can be handled by PT-MPO methods without further approximation.
This approach does not require expressions for the system eigenstates
and energies, and allows for genuine non-Markovian dynamics of many-body systems.
As we will discuss, mean-field theory consists of the ansatz that there are no
correlations between certain parts of the system. Here we employ this approach
to accurately treat the vibrational environments of a many-molecule--cavity
system.
\new{%
In particular, we develop a realistic model of an organic laser based on BODIPY-Br [\cref{fig:1a,fig:1b}], an organic molecule which has shown polariton lasing~\cite{Grant2016, Cookson2017}. We find results that differ significantly from those obtained in the model where the vibrational
environments cause simple dephasing---a model that cannot account for lasing in the presence of strong light-matter coupling. 
}
We determine how modifying the light-matter coupling and environmental temperature of our model changes the lasing threshold, and calculate the observed photoluminescence.

We model  \(N\) identical molecules as a collection of two-level systems
(Pauli matrices \(\sigma^\alpha_i\)) interacting with a single
near-resonant cavity mode (bosonic operator \(a^{\vphantom{\dagger}}_{}\))
according to the Dicke Hamiltonian under the rotating-wave approximation.
Setting \(\hbar=1\), the system Hamiltonian is
\begin{align}
	H_S = \omega_c a^{\dagger}_{}a^{\vphantom{\dagger}}_{} 
	+ \sum_{i=1}^N \left[\frac{\omega_0}{2} \sigma^z_i
	+  \frac{\Omega}{2\sqrt{N}} \left( a^{\vphantom{\dagger}}_{}
	\sigma^+_i + a^{\dagger}_{} \sigma^-_i \right)\right]
	\label{eq:HS}
\end{align}
where \(\omega_0\) and \(\omega_c\) are the two-level system and cavity
frequencies, and \(\sigma^+_i\) (\(\sigma^-_i\)) the raising (lowering)
operator for the \(i\)th spin.  The collective coupling \(\Omega\)
controls the light-matter interaction such that the bright eigenstates of
\(H_S\), i.e.\ the polaritons, are split as \(\pm\Omega/2\) at resonance.

The Hamiltonian \cref{eq:HS} may be referred to as the Tavis-Cummings model. Its
extension to include a single vibrational mode, the Holstein-Tavis-Cummings
model, has frequently been used to describe cavity bound organic
emitters~\cite{Cwik2014, Spano2015, Herrera2016, Wu2016,  Herrera2016a, Zeb2018,
Strashko2018, Arnardottir2020}. We instead consider the interaction of each
two-level system with a \textit{continuum} of modes represented by
the harmonic environment
\begin{align}
	H_{E}^{(i)} = \sum_{j} \left[
		\nu_{j} b^{\dagger}_{j} b^{\vphantom{\dagger}}_{j} 
	+ \frac{\xi_{j}}{2} 
	(b^{\vphantom{\dagger}}_{j}+b^{\dagger}_{j})\sigma^z_i\right]\text{,}
	\label{eq:HEi}
\end{align}
where \(b^{\vphantom{\dagger}}_{j}\) is the annihilation operator for the
\(j\)th mode of frequency \(\nu_j\).  The system-environment coupling
is characterized by a spectral density  \(J(\nu)= \sum_{j}  (\xi_j/2)^2
\delta(\nu-\nu_j)\), taken to be Ohmic in the form
\begin{align}
	J(\nu) = 2\alpha \nu e^{-(\nu/\nu_c)^2}\text{,} \quad \nu>0\text{,} 
	\label{eq:J}
\end{align}
where \(\alpha\) and \(\nu_c\) are chosen to reproduce the leading structure of
the absorption spectrum of BODIPY-Br at \(T=300~\text{K}\) [\cref{fig:1b}].
This effectively captures the low frequency modes arising from the host matrix
of the molecule.  The realistic picture of vibrational dephasing it affords is
the most significant advancement of our work.  In the limit that the
	system-environment coupling is weak one might look to derive a Redfield
	theory~\cite{Breuer2002}.  However, as we discuss in the Supplemental
Material~\cite{supplement}, this is difficult in the presence of strong
light-matter coupling.

Finally we consider incoherent pump \(\Gamma_\uparrow\) and dissipation
\(\Gamma_\downarrow\) of the two-level systems as well as  field decay
\(\kappa\).  Since these are associated with baths at optical frequencies (e.g.\
\(10^{15}\)~Hz) they may be well approximated~\cite{Breuer2002} by Markovian
terms in the master equation for the total density operator \(\rho\),
\begin{align}
	\begin{split}
		\partial_t \rho = &-i \biggl[ H_S + \sum_{i=1}^N H_E^{(i)}, \rho \biggr]
	+ 2 \kappa \mathcal{L}[a^{\vphantom{\dagger}}_{}]\\
	&+ \sum_{i=1}^N (\Gamma_\uparrow \mathcal{L}[\sigma^+_i]
	+  \Gamma_\downarrow \mathcal{L}[\sigma^-_i])\text{,}
\end{split}
	\label{eq:ME}
\end{align}
with \(\mathcal{L}[x]=x\rho x^\dagger - \{x^\dagger x, \rho\}/2\). 
\new{%
If \(H_E^{(i)}\) is absent one recovers the Tavis-Cummings model 
with pumping and decay which, as we discuss below, requires inversion \(\Gamma_\uparrow
> \Gamma_\downarrow\) to show lasing.
}
Below we fix
\(\Gamma_\downarrow\) and \(\kappa\) and observe the transition of the system
from a normal state, where the expectation \(\langle a^{\vphantom{\dagger}}_{}
\rangle\) of the photon operator vanishes, to a lasing state, where  \(\langle
a^{\vphantom{\dagger}}_{} \rangle\) is nonzero and time dependent, as
\(\Gamma_\uparrow\) is increased from zero.

Simulating dynamics in the presence of strong coupling to a structured
environment is a computationally intense task and as such PT-MPO methods cannot
be used to solve for a  large number of open systems simultaneously. Our
strategy is to use mean-field theory to reduce the \(N\)-molecule--cavity system
to a single molecule interacting with a coherent field [\cref{fig:1c,fig:1d}].

According to mean-field theory, we assume a product state for the many-body
density operator \(\rho\), i.e.\ a factorization between the photon and 
individual molecules, an ansatz known \cite{Mori2013,Carollo2021} to be exact as
\(N\to\infty\). This reduces the problem to the coupled dynamics of the
molecular mean-field Hamiltonian
\begin{align}
	H_{\text{MF}} = \frac{\omega_0}{2} \sigma^z +
	\frac{\Omega}{2\sqrt{N}}(\langle a^{\vphantom{\dagger}}_{} \rangle 
	\sigma^+ 	+ \langle a^{\vphantom{\dagger}}_{} \rangle^* \sigma^-), 
	\label{eq:HMF}
\end{align}
combined with evolution of the field expectation
\begin{align}
	\partial_t \langle a^{\vphantom{\dagger}}_{} \rangle
	&= -(i\omega_c+\kappa) \langle a^{\vphantom{\dagger}}_{}\rangle
	-i\frac{\Omega\sqrt{N}}{2} \langle \sigma^- \rangle \text{.}
	\label{eq:a} 
\end{align}
Here \(\langle \sigma^- \rangle\) (no subscript)
is the average of any one of the identical spins.
Thus, by propagating a \textit{single} spin with \(H_{\text{MF}}\) and subject
to the vibrational environment and individual losses described above, we can
effectively simulate the \(N\)-molecule system using a PT-MPO method provided
that at each time step  we also evolve \(\langle
a^{\vphantom{\dagger}}\rangle\) according to \cref{eq:a} [\cref{fig:1d}].
In Ref.~\cite{supplement} we discuss the derivation of \cref{eq:a,eq:HMF}
further as well as the role of ``bright'' and ``dark'' excitonic
states~\cite{houdr1996,Eastham2001,Herrera2016a,Herrera2016b,Cwik2016,Herrera2018}
in mean-field theory.

To calculate the dynamics  we use the PT-MPO provided by the
time evolving MPO (TEMPO) 
method~\cite{Strathearn2018, Strathearn2020,Fux2021,TimeEvolvingMPO}. 
Notable to our problem is that the system propagators depend on the field
\(\langle a^{\vphantom{\dagger}}_{} \rangle\), which depends self-consistently
on the state of the system.  A second-order Runge-Kutta method is used to
integrate the field from \(t_n\) to \(t_{n+1}\) whence it may be used in the
construction of the system propagator for the next time step.  Further
implementation details are provided in Ref.~\cite{supplement}.  Importantly the
construction of the PT capturing the influence of the bath, which is the costly
part of the calculation, only needs to performed once for a given spectral
density \cref{eq:J} and bath temperature \(T\).  It can then be reused with many
different system Hamiltonians or parameters.  This is particularly advantageous
when one wishes to vary one or more system parameters to map out a phase
diagram.

\begin{figure}
	\centering
    \vspace{-2\baselineskip}%
	\phantomsubfloat{\label{fig:2a}}%
    \phantomsubfloat{\label{fig:2b}}%
    \phantomsubfloat{\label{fig:2c}}%
    \phantomsubfloat{\label{fig:2d}}%

	\includegraphics[scale=0.395]{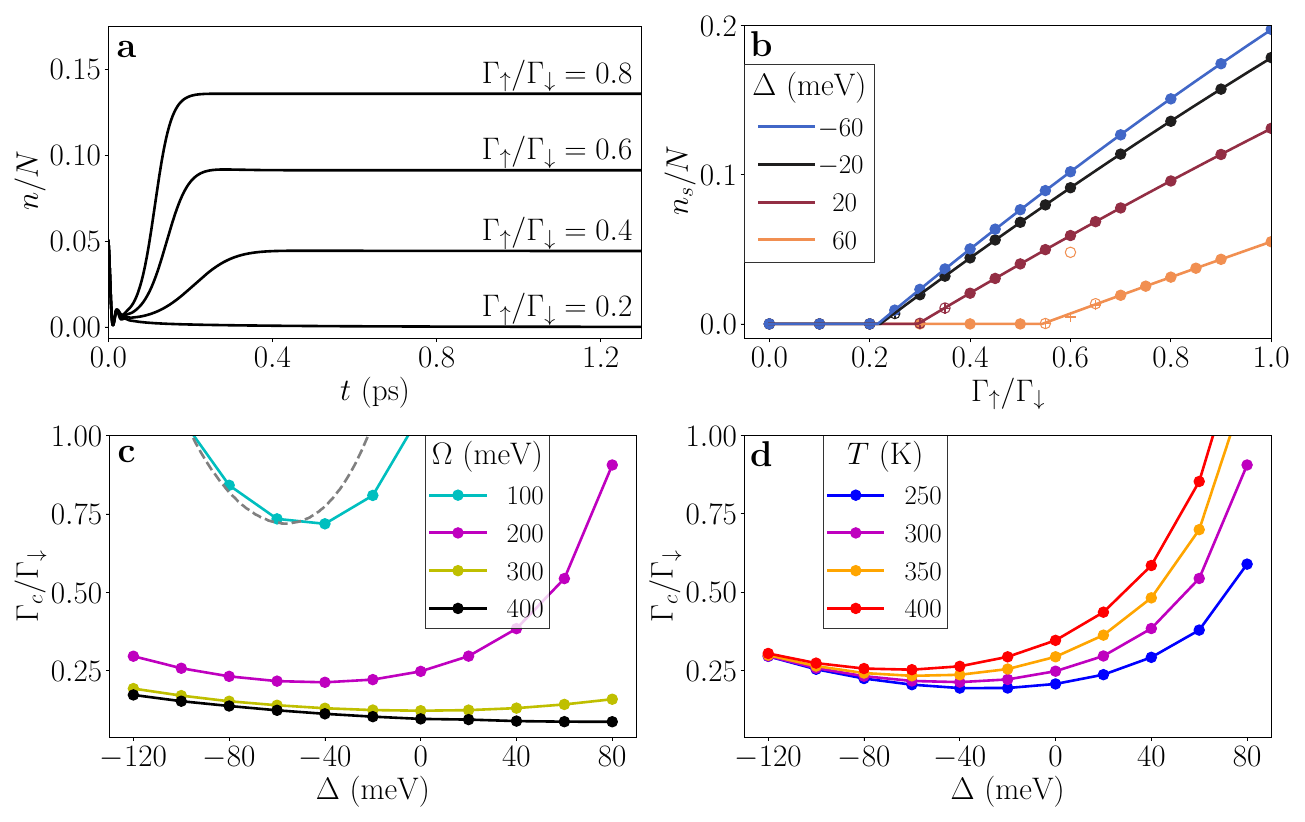}%
	\caption{%
		Determining the threshold of an organic laser. 
		(a) Example dynamics of the scaled photon number \(n/N=|\langle
		a\rangle|^2/N\) below (\(\Gamma_\uparrow = 0.2 \Gamma_\downarrow\)) and
		above (\(\Gamma_\uparrow \geq 0.4 \Gamma_\downarrow\)) the lasing
		transition at \(\Omega=200\)~meV, \(T=300\)~K and \(\Delta
		= \omega_c-\omega_0=-20\)~meV (note \(n\) scales with \(N\) above
		threshold~\cite{Kirton2019}).  The losses were fixed at
		\(\kappa=\Gamma_\downarrow=10\)~meV. Initial conditions: the
		system was prepared in a \(\sigma^z\)-down state with \(n_0/N=0.05\) and
		the bath in a thermal state.  Computational parameters and convergence
		information are provided in Ref.~\cite{supplement}.
		(b) Steady-state photon number with pump strength at \(\Omega=200\)~meV,
		\(T=300\)~K for several different detunings (closed circle: steady-state
		value obtained from a valid fit of late time behavior, open circle:
		invalid fit, cross: final value).  Fitting a curve to the data at each
		detuning provided an estimate of the threshold \(\Gamma_c\).  This was
		repeated for different \(\Omega\) and \(T\) to produce (c) and (d),
		respectively.  The result of a weak-coupling
		calculation~\cite{supplement} for \(\Omega=100\)~meV  is included in (c)
		as a dashed gray line.
	}
	\label{fig:2}
\end{figure}

\Cref{fig:2a} shows time evolution simulations at \(\Omega=200\)~meV and a small
negative detuning \(\Delta=\omega_c-\omega_0=-20\)~meV. For each run the bath
was prepared in a thermal state at \(T=300\)~K and the spin pointing down, with
a small initial field to avoid
the trivial fixed point of \cref{eq:a,eq:HMF}. The dynamics were generated up to
a time \(t_f=1.3\)~ps and the final value  \(\langle
a^{\vphantom{\dagger}}\rangle_f\) recorded.  This gave the steady-state field
except near the phase boundary where, due to the critical slowing down
associated with a second-order transition, \(\langle
a^{\vphantom{\dagger}}\rangle\) was still changing at \(t_f\).  To accommodate
this, an exponential fit was made to the late time dynamics yielding an estimate
of the steady-state value indicated by filled circles in \cref{fig:2b}.  Where
this was not possible (i.e.\ the fitting failed), the final value of the field
is marked with a cross and the attempted fit with an open circle.  An automated
procedure~\cite{supplement} was used to assess fit validity and any point with
an invalid fit was not used in subsequent calculations. 

Having obtained the steady-state field for a number of pump strengths
encompassing the transition [\cref{fig:2b}], a second fitting was performed to
extract the threshold pump \(\Gamma_c\)  at each detuning. This was repeated for
different light-matter coupling strengths and temperatures to produce the phase
diagrams \cref{fig:2c,fig:2d}.

In \cref{fig:2c}, we study the evolution of the threshold \(\Gamma_c\) as the
coupling \(\Omega\) increases.  At the smallest coupling con\-sidered, 
\(\Omega=100\)~meV, the  threshold is high and for
\(\Gamma_\uparrow\leq\Gamma_\downarrow\) there is only a small window of
detunings for which lasing is supported---i.e.,  the photon frequency coincides
with a region of net gain in the spectrum~\cite{schafer1990}. This curve may be
compared with the prediction of weak light-matter coupling
theory~\cite{supplement} shown with a gray dashed line. The disagreement here,
most apparent nearer zero detuning, reflects the fact that \(\Omega=100\)~meV is
already beyond weak light-matter coupling. 

We note the observed behavior cannot be described by a weak system-bath coupling
model in which the coupling to the bath is replaced by Markovian (temperature
dependent) dephasing.  Indeed, such a model requires
\(\Gamma_\uparrow>\Gamma_\downarrow\) for lasing and predicts a phase diagram
that is symmetric about \(\Delta=0\)~\cite{supplement}.  The same is
true for models that completely neglect the effect of vibrational
modes~\cite{Kirton2018}.  The existence of lasing for $\Gamma_\uparrow
< \Gamma_\downarrow$ within our model is a consequence of the vibrational bath.
The detuning for minimum threshold evolves with \(\Omega\) and is
not simply set by the peak of the  molecular emission spectrum; this is due to
reabsorption of cavity light playing a role for the parameters we
consider~\cite{Kirton2015}.

As the light-matter coupling increases, faster emission into the cavity mode
sees the threshold reduce before eventually saturating.  The
threshold becomes less dependent on detuning as lasing is now dictated by
whether the frequency of the lower polariton formed coincides with a region of
gain in the spectrum, and this occurs for a larger range of cavity frequencies.
Similar observations were made in models with sharp vibrational
resonances~\cite{Strashko2018}.  In that work reentrance under
\(\Gamma_\uparrow\) was seen---behavior absent here because of the broader
molecular spectrum we consider.

A key question in the study of organic polaritons is to what extent
thermalization occurs, and thus how temperature affects the
threshold~\cite{Kirton2015,Keeling2020}.  Motivated by this and the range of
temperatures accessible in organic polariton experiments we examine the dependence
of threshold on environmental temperature \(T\) at fixed \(\Omega=200\)~meV.
Changing \(T\) shifts, and increases the width of, the molecular spectrum. The
result for the phase diagram, shown in \cref{fig:2d}, is a suppression of lasing
with increasing \(T\), most significantly for positive detunings where the lower
polariton is more excitonic. This temperature dependence is one aspect of the
phase diagram that cannot generally be captured by simplified models with a few
vibrational modes, as we demonstrate in Ref.~\cite{supplement}.

We next study quadratic fluctuations about the mean field, as described by
two-time correlations and their Fourier transforms.  Specifically we calculate
the spectral weight and the photoluminescence (PL) spectrum, the latter of which
is the actual measured observable in all polariton
experiments~\cite{Keeling2020}.  Multitime correlations are naturally
accessible within the PT-MPO framework, allowing us to calculate absorption and
emission spectra without recourse to the quantum regression theorem. 

We use that the retarded \(D^R\) and Keldysh \(D^K\) photon Green's functions
may be written in terms of the exciton self-energies~\cite{Cwik2016,
Keeling2010}
\begin{align}
\Sigma^{-+}(\omega) &= \frac{i\Omega^2}{4}\int_{0}^\infty dt e^{i\omega t} 
\langle [\sigma^-(t),\sigma^+(0)] \rangle\text{,}
\label{eq:smp}\\
\Sigma^{--}(\omega) &= \frac{i\Omega^2}{4}\int_{-\infty}^\infty dt e^{i\omega t}
\langle \{\sigma^-(t),\sigma^+(0)\} \rangle\text{.}\label{eq:smm}
\end{align}
The photon Green's functions then take the form 
\begin{align}
D^R(\omega) &= \frac{1}{\omega-\omega_c+i\kappa+ 
\Sigma^{-+}(\omega)}\label{eq:DR}\text{,}
	\\
	D^K(\omega)
	&= -\frac{\Sigma^{--}(\omega)+ 2i\kappa}{\abs{\omega-\omega_c+i\kappa+
	\Sigma^{-+}(\omega)}^2}\text{.}\label{eq:DK}
\end{align}
Hence, by calculating the correlators \(\langle \sigma^-(t)\sigma^+(0)\rangle\)
and \(\langle \sigma^+(t)\sigma^-(0)\rangle\) using the PT-MPO approach, we can
find the Green's functions \(D^R\) and \(D^K\) which fully characterize
the spectrum of the nonequilibrium system.

Thus far we have considered a model with a single photon mode for which
	mean-field theory is exact as \(N\to\infty\).  However, it is
	straightforward to extend our analysis to include multiple photon modes, where
	mean field can still provide a good approximation~\cite{supplement}.  Hence
	we consider the model with cavity mode term \( \sum_{\vect{k}}
	\omega_{c,\vect{k}} a^{\dagger}_{k} a^{}_{\vect{k}}\), where \(\omega_{c, k}
	= \omega_c + k^2/2m_{\text{ph}}\) (recall \(\hbar=1\)), and light-matter
	interaction \(\sum_{\vect{k}} \Omega a^\dagger_{\vect{k}}e^{i
	\vect{k}\cdot \vect{r}_n} \sigma_i^- +\text{H.c.}\). As
	discussed in Ref.~\cite{supplement}, the mean field steady-state equations
	remain similar and one now has access to the photon Green's functions
\(D^R_{\vect{k}}(\omega)\), \(D^K_{\vect{k}}(\omega)\) of the	multimode
model.

\begin{figure}
	\centering
    \vspace{-2\baselineskip}%
	\phantomsubfloat{\label{fig:3a}}
    \phantomsubfloat{\label{fig:3b}}
    \phantomsubfloat{\label{fig:3c}}
    \phantomsubfloat{\label{fig:3d}}

	\includegraphics[scale=0.395]{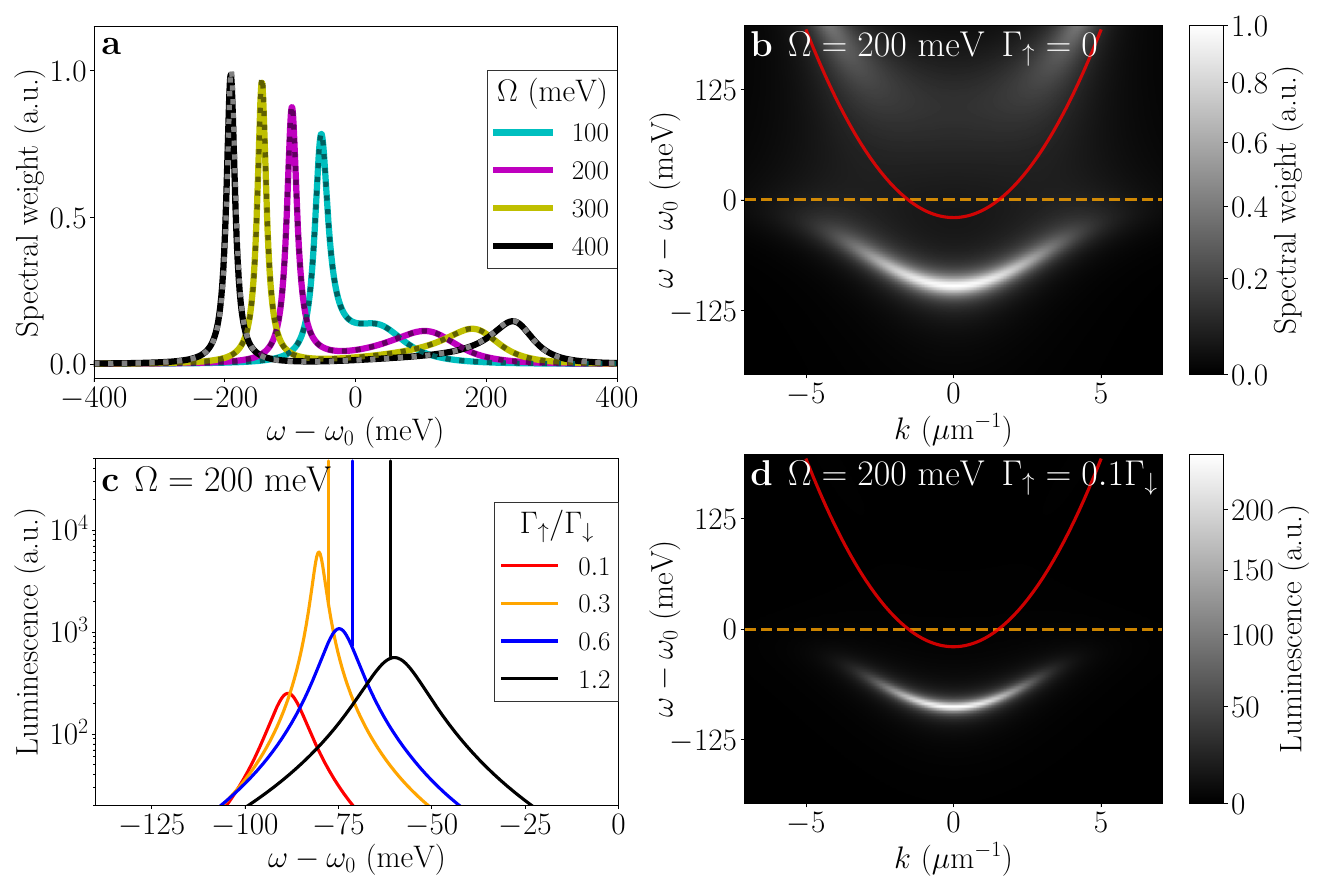}
	\caption{%
		(a) Spectral weight, \cref{eq:absorption}, at \(k=0\) when
		\(\Gamma_\uparrow=0\).  At each light-matter coupling, results from the
		analytic self-energy are shown as a dotted line, and results from PT-MPO
		as a solid line.
		(b) \(k\)-dependent spectral weight for \(\Omega=200\)~meV. The bare
		molecular energy \(\omega_0\) is shown in orange and the photon
		dispersion \(\omega_{c, k}\) in red (the photon mass \(m_{\text{ph}}\)
		was set to \(\omega_c/c^2\)).
		(c) Photoluminescence, \cref{eq:pl}, at \(k=0\) on a logarithmic scale
		for four different pump strengths at \(\Omega=200\)~meV. Above threshold
		the spin-spin correlators have a nonzero long time value giving
		a delta singularity i.e.\ lasing peak in the spectrum, indicated here as
		a vertical line.  Additional cross sections at smaller and larger
		\(\Omega\) are provided in Ref.~\cite{supplement}.
		(d) $k$-dependent photoluminescence below threshold at
		\(\Omega=200\)~meV and \(\Gamma_\uparrow=0.1\Gamma_\downarrow\) with red
		and orange lines as in (b).  All panels were produced at \(\Delta
		= -20\)~meV and \(T=300\)~K, with losses
		\(\kappa=\Gamma_\downarrow=10\)~meV. 
}
	\label{fig:3}
\end{figure}

We first consider the system without pumping (\(\Gamma_\uparrow=0\)) and the
spectral weight~\footnote{We denote this as spectral weight rather than
absorption, as the absorption spectral of a general lossy cavity is a more
complicated expression, see Refs.~\cite{Ciuti2006IO,Cwik2016} for discussion.}  
\begin{align}
	\rhospec_{\vect{k}}(\omega) = - 2 \text{Im}D^R_{\vect{k}}(\omega) \text{.} 
	\label{eq:absorption}
\end{align}
As the system is in the normal state, 
\(\langle \sigma^+(t)\sigma^-(0)\rangle\equiv0\),
while an exact expression for the other correlator may be
found~\cite{Kirton2015} as \(\langle \sigma^-(t)\sigma^+(0) \rangle
= e^{-i\omega_0 t - \phi(t) - (\Gamma_\downarrow/2)t}\) where
\begin{equation}
\phi(t) = 
\int_{-\infty}^\infty d\omega \frac{J(\omega)}{\omega^2}
\left[2\coth\left(\frac{\omega}{2T}\right)
\sin^2\left( \frac{\omega t}{2} \right) + i \sin(\omega t)\right]
\text{.}
\label{eq:C}
\end{equation}
This provides a benchmark of our numerics:
\Cref{fig:3a} shows excellent agreement between the  spectral
	weight derived from the analytical result \cref{eq:C} and that from
	measurement of the correlator using the PT-MPO method at \(k=0\).
	Figure~\ref{fig:3b} then illustrates the \(k\) dependence of the spectrum for
\(\Omega=200\)~meV. 

When the system is pumped, i.e.\ 
\(\Gamma_\uparrow \neq 0\),  no  analytical
results are available and it is necessary to determine both the spectrum and its
occupation numerically. Here we calculate the
photoluminescence~\cite{Keeling2010},
\begin{align}
	\mathcal{L}_{\vect{k}}(\omega) = \frac{i}{2} 
	\left( D^K_{\vect{k}}(\omega)- 
	D^R_{\vect{k}}(\omega) 
+ [D^R_{\vect{k}}(\omega)]^\ast \right)\text{.}
	\label{eq:pl}
\end{align}
Figure~\ref{fig:3c} shows \(\mathcal{L}_{\vect{k}=0}(\omega)\) at fixed
	detuning \(\Delta=-20\)~meV and \(\Omega=200\)~meV for four different pump
	strengths.  At the weakest pump strength,
	\(\Gamma_\uparrow=0.1\Gamma_\downarrow\), the system is below threshold yet
	\(\mathcal{L}_{\vect{k}}(\omega)\) does not vanish since, in contrast to the
	mean-field calculation of the steady-state photon number, the
	photoluminescence contains an incoherent part. Plotting the \(k\) dependence
of the spectrum in this case [\cref{fig:3d}] makes clear this arises from the
lower polariton.  

At higher pump strengths, \(\Gamma_\uparrow=0.3, 0.6,
	1.2\Gamma_\downarrow\) in \cref{fig:3c}, the system is above threshold, with
	the coherent lasing contribution indicated by a delta peak superimposed on
	the spectrum.  In particular, for \(\Gamma_\uparrow=0.3\Gamma_\downarrow\)
	and \(0.6\Gamma_\downarrow\), the lasing frequency occurs noticeably to the
	right of the peak luminescence: the conditions to maximize
	\(\mathcal{L}_{\vect{k}}\), which depends on both the density of states and
	their populations, do not, in general, coincide with the point at which the
	lasing instability develops.  We explore this further in
	Ref.~\cite{supplement} by examining the real and imaginary parts of the
	inverse Green's functions as the transition is approached. 

In conclusion, we have developed a technique for calculating the
	non-Markovian dynamics of a many-body open system using mean-field theory
	and PT-MPO methods. We applied this technique to model the polariton lasing
	of an organic dye in a microcavity including many molecules with realistic
	vibrational spectra. This provided the steady-state of the
	driven-dissipative system and, via the measurement of two-time correlations,
	its spectrum. We first determined the dependence of the threshold for lasing
	on cavity detuning under different light-matter coupling strengths and
	environmental temperatures.  Second, we observed how the photoluminescence
	and lasing frequency of the model evolved with pump strength.  For the case
	of a one-to-all interaction between the cavity and molecules, the mean-field
	treatment is exact as $N\to\infty$~\cite{Mori2013,Carollo2021}.  The same
	applies to all-to-all networks of open systems~\cite{Carollo2021,supplement}.  More
	generally, there are situations  where mean-field theory is not exact but
offers a good approximation, including models of polariton condensation with
multiple modes such as considered in Refs.~\cite{keeling2004,Arnardottir2020,supplement}.

\acknowledgments
We thank  G. E. Fux and P. Kirton from the TEMPO Collaboration for helpful
discussions when implementing the mean-field approach.  
P.F.-W. acknowledges support from EPSRC (EP/T518062/1).
B.W.L. and J.K. acknowledge support from EPSRC (EP/T014032/1).

\bibliography{refs.bib}

\clearpage

\renewcommand{\theequation}{S\arabic{equation}}
\renewcommand{\thefigure}{S\arabic{figure}}
\setcounter{equation}{0}
\setcounter{figure}{0}
\setcounter{page}{1}

\onecolumngrid
\titleformat{\section}
{\small\bfseries\filcenter}{\thesection}{1em}{\MakeUppercase{#1}}
\section{Supplementary material for: Efficient many-body non-Markovian dynamics of organic polaritons}
\twocolumngrid
\renewcommand{\thesection}{\arabic{section}}
\renewcommand{\thesubsection}{\arabic{subsection}}
\titleformat{\section}
{\small\bfseries\filcenter}{\value{section}}{1em}{\stepcounter{section}\hypertarget{sm:\thesection}{\thesection.~\MakeUppercase{#1}}}
\setcounter{section}{0}

\section{Weak system-environment coupling}%
In the limit where the system-environment coupling is sufficiently weak, one
might expect it would be possible to derive and use an accurate time-local
(Markovian) description.  In this section we discuss the challenges in doing
this and explain why, even in this weak system-bath coupling limit, the PT-MPO
approach may still be valuable.

When the system-environment coupling is weak, one can apply standard
methods~\cite{Breuer2002} to derive a Redfield theory describing the low
frequency vibrational environment.  In appropriate cases, one can further
secularize this Redfield theory to give a density matrix equation of motion of
the Gorini–Kossakowski–Sudarshan–Lindblad form~\cite{Breuer2002}.  For our
model, considering system-bath coupling as written in \cref{eq:HEi}, the
contribution to the density matrix equation describing this bath takes the form:
\begin{equation}
   \left.\partial_t\rho\right|_{\text{vib.}} = 
   \sum_{i,n} \left[ \Gamma(\lambda_n)
   \left(
      \varsigma^z_{i,n} \rho \sigma^z_i - \sigma^z_i \varsigma^z_{i,n} \rho
   \right) + \text{H.c.}
   \right].
\end{equation}
 Here $\Gamma(\lambda)
= \int_0^\infty \!\!ds e^{i \lambda s} C(s)$ where $C(s)$ describes correlations
of the bath operators which couple to the system, ${x}_j = {b^{}_j
+ b_j^\dagger}$:
\begin{align}
    C(s)&\equiv
    \sum_j \left(\frac{\xi_j}{2}\right)^2
    \langle {x}_j(t) {x}_j(t-s)
    \rangle
    \nonumber\\
    &=
    \int d \nu J(\nu) \left[ \coth\left(\frac{\nu}{2 T} \right) 
	\cos(\nu s) - i \sin(\nu s) \right].
\end{align}
The operators $\varsigma^z_{i,n}$ are the eigen-operator decomposition of
$\sigma^z_i$.  They obey $[H_S,\varsigma^z_{i,n}]=-\lambda_n \varsigma^z_{i,n}$
where $H_S$ is the system Hamiltonian, and satisfy $\sum_n
\varsigma^z_{i,n}=\sigma^z_i$.  Formally they can be found using the eigenstates
of $H_S \ket{n}= \epsilon_n \ket{n}$, by writing a restricted sum over
transitions with energy difference $\lambda_n$
\begin{equation}
    \varsigma^z_{i,n} = 
    \sum_{\substack{m,p\\\epsilon_m=\epsilon_p-\lambda_n}} 
    \ket{m}\!\!\mel{m}{\sigma^z_i}{p}\!\bra{p}.
\end{equation}
Evaluating this however presents a severe problem for the Tavis--Cummings model
with strong light-matter coupling, as it requires expressions for the complete
spectrum of eigenstates and energies.  In general, for many-body problems, this
is not available.

There do exist some special cases where one can give explicit forms of the
dissipation.   The simplest case---which recovers the phenomenological picture
of vibrations causing dephasing---is to neglect light-matter coupling in
deriving $\varsigma^z_{i,n}, \lambda_n$.  In this case there is a single
eigen-operator $\varsigma^z_{i,0}=\sigma^z_i, \lambda_0=0$, and one finds a pure
dephasing process.  For the Ohmic spectrum $J(\nu)$ defined in the main text one
finds $4 \pi \alpha T \sum_i \mathcal{L}[\sigma^z_i]$.  The behavior of the
driven-dissipative Tavis--Cummings model with dephasing has been extensively
studied elsewhere (see e.g.\ Ref.~\cite{Kirton2019}).  In such a model lasing only
occurs for $\Gamma_\uparrow>\Gamma_\downarrow$, and the threshold ratio
$\Gamma_\uparrow/\Gamma_\downarrow$ is symmetric around cavity-molecule detuning
$\Delta=0$.  Both these features are notably different to the results seen in
Fig.~\ref{fig:2}.  We may also note that the same statements apply when there is
no effect of the vibrational bath at all. In that case our model becomes the
Tavis--Cummings model with only pumping $\Gamma_\uparrow$, and decay
$\Gamma_\downarrow, \kappa$ processes. As discussed extensively in previous
work, e.g.~\cite{haken1970,Kirton2018}, this model also requires
$\Gamma_\uparrow>\Gamma_\downarrow$ for lasing to occur.
\new{%
However, polariton splitting is suppressed at large pumps strengths,
so such models cannot provide a description of experiments~\cite{Kena2010,
Plumhof2014, Daskalakis2014, Grant2016, Dietrich2016,Cookson2017} 
demonstrating polariton lasing in the strong light-matter coupling regime.
}

Another case where explicit results can be derived is at weak excitation, when
the saturable two-level operators $\sigma^\pm_i$ can be replaced by bosonic
operators $c_i^\dagger,c_i$.  This yields a system Hamiltonian that is quadratic
in bosonic operators, and can be solved exactly, see
Refs.~\cite{pino2015,MartinezMartinez2018:Comment,Pino2018:Response}.  However,
neglecting saturation of the two-level system is not valid when considering
strong driving and lasing.

The fact that microscopic derivation of dissipation requires knowledge of the
eigenspectrum of the system Hamiltonian in fact provides further motivation for
methods such as the mean-field PT-MPO approach.  That is, even when a weak
coupling approach might be valid, it may not always be practical to evaluate the
eigen-operators and values. Approaches based on the PT-MPO remove this
requirement, enabling one to study the dynamics of many-body systems coupled to
structured environments.

\section{Mean-field Equations}%
In this section we derive the mean-field Hamiltonian \cref{eq:HMF} and equation
of motion \cref{eq:a}.  As noted in the Letter, for models with many-to-one
coupling, such as our emitter--cavity model, it can be shown
\cite{Mori2013,Carollo2021} that a mean-field ansatz is exact as \(N\to
\infty\). 

In its most general form, mean-field theory can be understood as an assumption
about the structure of the state of a many-body system
\cite{chaikin1995,wipf2021,kramer2015}.  Specifically, for our model, this means
to consider the product state
\begin{align}
	\rho = \rho_{a} \otimes \bigotimes_{i=1}^N \rho_i. \label{eq:rhomf}
\end{align}
The reduced density matrix \(\rho_a= \Tr_{\otimes{i}}\rho\) is obtained from the
partial trace taken over the Hilbert space of all two-level systems labelled
\(i=1,2,\ldots, N\), and \(\rho_i = \Tr_{a, \otimes{j\neq i}} \rho\) from the
partial trace over the photonic degree of freedom and all but the
\(i^{\text{th}}\) two-level system.

In the calculations presented in the main text, we make a further simplification
by taking all molecules to be identical, so that only a single $\rho_i$ needs to
be calculated.  We note however that the mean-field method we describe here does
not require this.  The mean-field treatment can be applied to models where each
molecular site has different parameters, at the cost of requiring separate
simulations for each $\rho_i$.   We also note that even when all sites are
equivalent, the assumption of identical $\rho_i$ is not the same as restriction
to the totally symmetric Hilbert space, particularly when incoherent processes
are present.  We discuss the consequences of this further below, in terms of the
role of ``dark'' exciton states within mean-field theory.

In our approach both the non-Markovian environment and  Markovian pumping and
loss for each molecule are handled by the PT-MPO method.  As discussed below,
the PT-MPO can be derived starting from the dynamics of the density matrix of an
individual molecule $\rho_i$.  That is, such dynamics could be considered as
part of the evolution of $\rho_i$, included within the system Hamiltonian, and
then handled through the PT-MPO approach.  However, explicitly including such
dynamics in our discussion of the mean-field approximation makes the derivation
appear unnecessarily complicated.  We therefore discuss the mean-field
decoupling approach to dynamics for a simpler model (the Tavis--Cummings model),
and then re-introduce molecular dissipation terms in \hyperlink{sm:4}{Sec.~4}.  As such
we start from the following master equation:
\begin{align}
	\partial_t \rho = -i[H_S,\rho] + 2\kappa \mathcal{L}[a^{\vphantom{\dagger}}_{}],
\end{align}
along with the system Hamiltonian from the Letter,
\begin{align}
	H_S = \omega_c a^{\dagger}_{}a^{\vphantom{\dagger}}_{} 
	+ \sum_{i=1}^N \left[\frac{\omega_0}{2} \sigma^z_i
	+  \frac{\Omega}{2\sqrt{N}} \left( a^{\vphantom{\dagger}}_{} \sigma^+_i 
+ a^{\dagger}_{} \sigma^-_i \right)\right].
	\tag{\ref{eq:HS}}
\end{align}

The equations of motion for the reduced density matrices follow from
\begin{align}
	\partial_t \rho_{a} &= -i \Tr_{\otimes i} [H_S,\rho] + 2\kappa\Tr_{\otimes i}
	\mathcal{L}[a^{\vphantom{\dagger}}_{}],\label{eq:rhoa}\\
	\partial_t \rho_{i} &= -i \Tr_{a,\otimes j\neq i} [H_S,\rho]
	+ 2\kappa\Tr_{a,\otimes j\neq i}	
	\mathcal{L}[a^{\vphantom{\dagger}}_{}].\label{eq:rhoi}
\end{align}

The partial traces can be performed by noting two points. First, the separate
reduced density matrices are normalized to one. Second, the partial trace over
subsystem $I$ of a commutator involving operators acting only on subsystem $I$
will vanish.  Thus,
\begin{align}
	-i\Tr_{\otimes i}  [\omega_c a^{\dagger}_{}a^{\vphantom{\dagger}}_{}, \rho]
	&=	-i\omega_c [a^{\dagger}_{}a^{\vphantom{\dagger}}_{}, \rho_a],\\
	-i\Tr_{\otimes j} 
	\biggl[\sum_{i=1}^N\frac{\omega_0}{2} \sigma_i^z, \rho\biggr] &= 0,\\
	2\kappa\Tr_{\otimes i}  
	\mathcal{L}[a] &= 2\kappa \mathcal{L}_a[a^{\vphantom{\dagger}}_{}],
\end{align}
 where \(\mathcal{L}_a[x] =x\rho_a x^\dagger - \{x^\dagger x, \rho_a\}/2\) is the Lindblad
 operator for the photon density matrix, and
 \begin{align}
		-i  \Tr_{a,\otimes j\neq i}[\omega_ca^{\dagger}_{} a^{\vphantom{\dagger}}_{},
		\rho] &=0, \\
		-i\Tr_{a,\otimes j\neq i} 
		\biggl[\sum_{k=1}^N\frac{\omega_0}{2} \sigma_k^z, \rho\biggr] &= 
	-i \left[\frac{\omega_0}{2}\sigma_i^z, \rho_i\right],\\
	2\kappa\Tr_{a,\otimes j\neq i}	\mathcal{L}[a^{\vphantom{\dagger}}_{}]&=0.
\end{align}

It remains to determine the terms arising from the light-matter interaction in
\(H_S\).  For the contribution to the evolution of the photon degree of freedom
\cref{eq:rhoa}, one has
\begin{align}
	\begin{gathered}
		-i\sum_{i=1}^N\frac{\Omega}{2\sqrt{N}}
		\left(\Tr_{\otimes j}  [a^{\vphantom{\dagger}}_{}\sigma_i^+, \rho]
		+ \text{H.c.}\right)\\
			 =-i \frac{\Omega \sqrt{N}}{2}  \left( \langle\sigma^+ \rangle
 [a^{\vphantom{\dagger}}_{},	 \rho_a]+ 
 \langle\sigma^- \rangle
 [a^{\dagger}_{},	 \rho_a]
 \right).
\end{gathered}
 \end{align}
For the evolution of the matter degree of freedom \cref{eq:rhoi}, the
contribution is instead
\begin{align}
	\begin{gathered}
		-i\sum_{k=1}^N\frac{\Omega}{2\sqrt{N}} 
		\left(
	\Tr_{a,\otimes j\neq i} [a^{\vphantom{\dagger}}_{}\sigma_k^+,
		\rho]
		+ \text{H.c.}
		\right)
		\\
 		=-i\frac{\Omega}{2\sqrt{N}} \left(
 		\langle a^{\vphantom{\dagger}}_{} \rangle  [\sigma^+_{i}, \rho_i]
 		+\langle a^{\dagger}_{} \rangle  [\sigma^-_{i}, \rho_i]
 		\right).
\end{gathered}
\end{align}

From the above we find that the equation of motion for each molecule \(\rho_i\)
is
\begin{align}
	\partial_t \rho_i &= -i [H_i, \rho_i], \label{eq:rhoi2}
\end{align}
where	
\begin{align}
	H_i =   \frac{\omega_0}{2} \sigma_i^z + \frac{\Omega}{2\sqrt{N}}(\langle a^{\vphantom{\dagger}}_{} \rangle 
	\sigma^+_i 	+ \langle a^{\vphantom{\dagger}}_{} \rangle^* \sigma^-_i).
	\label{eq:HMFi}
\end{align}
is the mean-field Hamiltonian \(H_{\text{MF}}\), \cref{eq:HMF}, for one of the
identical emitters. In the full dissipative model, $H_i$ would also include the
bath terms for that molecule, and could be used used to construct the system
propagators in the PT-MPO method described in \hyperlink{sm:4}{Sec.~4}.  One may
note that in \cref{eq:HMFi}, the only property of the photon state $\rho_a$
required is the expectation \(\langle a^{\vphantom{\dagger}}_{} \rangle\).  One
may thus take the equation of motion for \(\rho_a\),
\begin{align}
	\partial_t \rho_a &= -i [H_a, \rho_a] + 
	2\kappa \mathcal{L}_a [a^{\vphantom{\dagger}}_{}],\label{eq:rhoa2}
\end{align}
with the Hamiltonian 
\begin{align}
	H_a = \omega_c a^{\dagger}_{}a^{\vphantom{\dagger}}_{}
	+ \frac{\Omega \sqrt{N}}{2}
	\left( a^{\vphantom{\dagger}}_{} \langle \sigma^+ \rangle
	+ a^{\dagger}_{} \langle \sigma^- \rangle\right),
\end{align}
and derive the equation of motion for
\(\langle a^{\vphantom{\dagger}}_{} \rangle\):
\begin{align}
	\partial_t\langle a^{\vphantom{\dagger}}_{} \rangle 
	&=  \Tr_a\left(a^{\vphantom{\dagger}}_{} \partial_t\rho_a\right) \nonumber\\
	\begin{split}
		&= -i\omega_c \Tr_a\left( a^{\vphantom{\dagger}}_{}
		[a^{\dagger}_{}a^{\vphantom{\dagger}}_{}, \rho_a]\right)
	- i \frac{\Omega \sqrt{N}}{2} \langle \sigma^-_{} \rangle 
	\Tr_a \left(a^{\vphantom{\dagger}}_{} [a^{\dagger}_{}, \rho_a]\right)\notag{}\\
	&\phantom{=}+ 2\kappa \Tr_a\left(a^{\vphantom{\dagger}}_{} 
		a^{\vphantom{\dagger}}_{}\rho_a a^{\dagger}_{}
	- a^{\vphantom{\dagger}}_{}a^{\dagger}_{} a^{\vphantom{\dagger}}_{}\rho_a/2
- a^{\vphantom{\dagger}}_{}\rho_a a^{\dagger}_{}a^{\vphantom{\dagger}}_{}/2\right)
\label{eq:a2}
\end{split}\\
&= -(i\omega_c+\kappa) \langle a^{\vphantom{\dagger}}_{}\rangle
	-i\frac{\Omega\sqrt{N}}{2} \langle \sigma^- \rangle .
\end{align}

\subsection{Field rescaling}
In the lasing phase \(\langle a^{\vphantom{\dagger}}_{} \rangle\) scales with
\(\sqrt{N}\) so it is convenient to  work with the rescaled quantity \(
\langle\tilde{ a} \rangle = \langle a^{\vphantom{\dagger}}_{} \rangle/\sqrt{N}\)
such that \cref{eq:a,eq:HMF} become
\begin{align}
	\partial_t \langle \tilde{a} \rangle &= 
	- (i\omega_c+\kappa)\langle \tilde{a} \rangle- 
	i \frac{\Omega}{2}\langle\sigma^-\rangle
	\intertext{and}
	H_\text{MF} &= 
 \frac{\omega_0}{2}\sigma^z+
	\frac{\Omega}{2}\left( \langle \tilde{a} \rangle \sigma^+ +
	\langle \tilde{a} \rangle^{*}\sigma^- \right)\text{.}  \label{eq:HS_rescale}
\end{align}
Hence only a single parameter \(\Omega\) is used to specify the light-matter
interaction.  It is the rescaled photon number, \(|\langle \tilde{a} \rangle|^2
\equiv n/N\), that is plotted in \cref{fig:2a,fig:2b}.

\section{Bright and dark exciton states in mean-field theory}

In this section we discuss the role that bright and dark excitonic states play
within a mean-field approach.  As discussed
elsewhere~\cite{houdr1996,Eastham2001,Cwik2016,ribeiro2018}, for a model of $N$
molecules coupled to a single photon mode, one can divide excitons into a single
optically ``bright'' mode---the spatially uniform superposition which couples to
the cavity mode, and $N-1$ ``dark'' modes which are orthogonal to the bright
mode.  The bright modes hybridize with the cavity mode to form polaritons, while
the dark modes remain at the bare exciton energy~\footnote{Extensions of this
concept can also be made for models including a continuum of in-plane cavity
modes; a similar division survives as long as the number of low energy photon
modes is much smaller than the number of
molecules~\cite{Agranovich2003,michetti2005polariton,martinez2019triplet,Keeling2020}.}.

When the molecules are disordered (e.g.\ different on-site energies), this mixes
the bright and dark states~\cite{houdr1996}, leading to a non-vanishing spectral
weight from the dark modes.  Since our model has no disorder, one might expect
the dark modes are absent.  However, as we discuss here, one can directly show
that within a mean field treatment, both bright and dark states are occupied.
Furthermore, despite the absence of static disorder, the vibrational environment
provides a form of dynamical disorder which makes the dark modes optically
active~\cite{Herrera2016a,Herrera2016b,Cwik2016,Herrera2018}.

\subsection{Exciton populations}

We first show how one can extract exciton populations from the mean-field
theory, and show that both the \(k=0\) ``bright'' states, as well as the
\(k\neq0\) ``dark'' states are populated.  

\begin{figure}
	\centering
	\vspace{-2\baselineskip}%
	\phantomsubfloat{\label{fig:sm_populationa}}
	\phantomsubfloat{\label{fig:sm_populationb}}
	
	\includegraphics[scale=0.395]{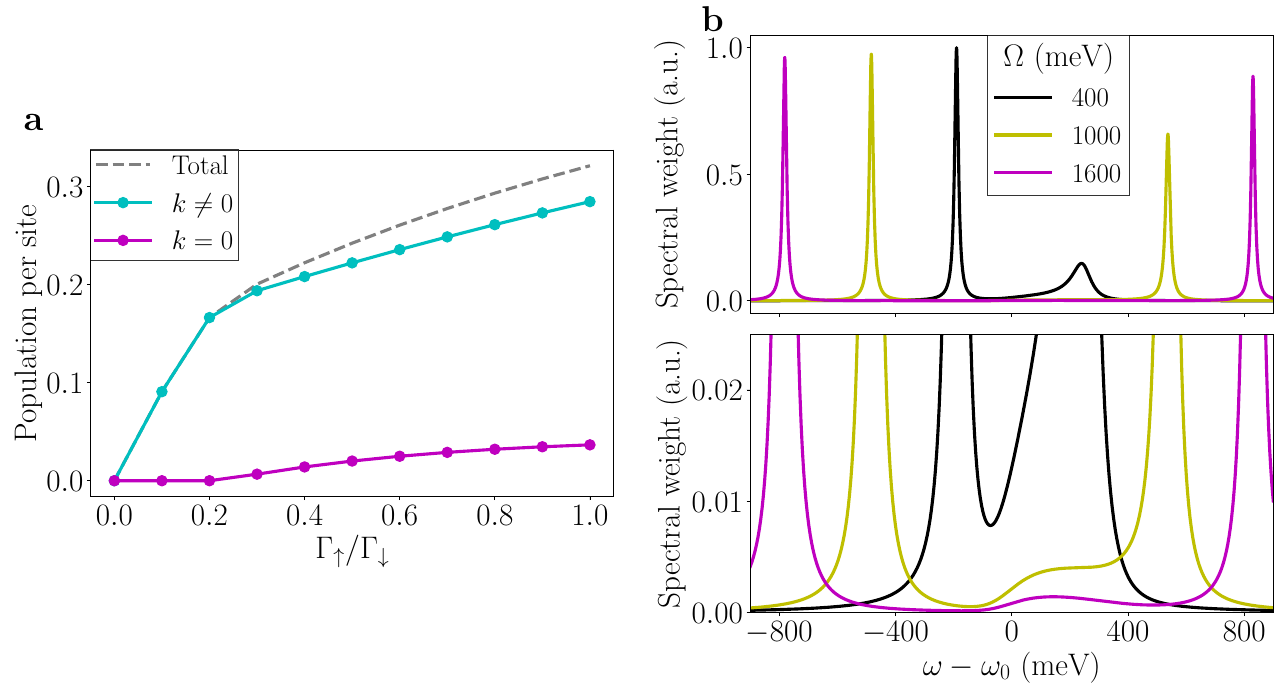}
	\caption{%
		(a) Exciton populations per site in the steady-state obtained using the
		PT-MPO method at \(\Omega=200\)~meV.  Below threshold the population per
		site of the \(k=0\) mode (or any single mode) vanishes as $1/N$.  The
		\(k=0\) population becomes macroscopic above threshold.  (b) Absorption
		spectrum, showing the existence of a residual excitonic peak at
		\(\Omega=1000\), \(1600\)~meV. Both panels show the same data on
		different vertical scales.  No residual peak is seen in the curve at
		\(\Omega=400\)~meV, which is the largest light-matter coupling strength
		considered in the main text. This is due to the proximity of the upper polariton
		whose tail swamps the residual peak.  Note the frequency structure of the
		vibrational environment means that this feature occurs at frequencies just above
		the zero-phonon line \(\omega=\omega_0\).  In this figure, the values of
		other parameters match those used in \cref{fig:2a} of the Letter
		(\(\Delta=-20\)~meV, \(T=300\)~K, \(\kappa=\Gamma_\downarrow=10\)~meV).
	}
	\label{fig:sm_population}
\end{figure}

Firstly, the total exciton population is:
\begin{align}
	P_{\text{tot.}} &= \sum_{i=1}^N \langle \sigma^+_{i}\sigma^-_{i} \rangle
	=\frac{N}{2} \left( 1+ \langle \sigma^z_{} \rangle \right)\label{eq:ptot}
\end{align}
where we write \(\langle \sigma^z_{}\rangle\) for the expectation at any one of
the \(N\) identical sites.  To find the bright and dark state populations, we
can consider exciton modes with defined momenta corresponding to creation
operators $\sum_i \sigma_i^+ e^{-i \vect{k} \cdot \vect{r}_i}/\sqrt{N}$.
Following this, the \(k=0\) exciton population is defined as
\begin{equation}
P_{\vect{k}=0} = \frac{1}{N} \sum_{i,j=1}^N \langle \sigma^+_{i} \sigma^-_{j} \rangle.
\end{equation}
Using the mean-field decoupling \( \langle \sigma^+_{i} \sigma^-_{j} \rangle=
\langle \sigma^+_{i} \rangle \langle \sigma^-_{j} \rangle\) for distinct sites
$i \neq j$ and the properties of Pauli operators for $i=j$,  the \(k=0\)
(bright) population is readily calculated as
\begin{align}
	P_{\vect{k}=0} 
	&= \frac{1}{N} \sum_{i=1}^N \frac{1}{2}
	\left( 1+ \langle \sigma^z_{} \rangle \right)
	+ \frac{1}{N} \sum_{j\neq i} \langle \sigma^+_{i}
	\rangle \langle \sigma^-_{j} \rangle\notag{} \\
	&= \frac{1}{2} \left( 1+ \langle \sigma^z_{} \rangle \right)
	+ (N-1) \abs{\langle \sigma^+_{} \rangle}^2 \notag{}.
\end{align}
By completeness of any $k$-space representation, the total population of dark
states can then be found as 	$P_{\vect{k}\neq0}
= P_{\text{tot.}}-P_{\vect{k}=0}$.  Since $P_{\vect{k}=0} \neq P_{\text{tot.}}$
one may clearly see that the mean-field approximation does not neglect the dark
state population.  The expressions for bright and dark mode populations simplify
when we consider the limit of large \(N\). In this case we may write:
\begin{align}
P_{\vect{k}=0} &\simeq N 	\abs{\langle \sigma^+_{} \rangle}^2,\label{eq:p0}
\\
P_{\vect{k}\neq0} &\simeq \frac{N}{2} \left( 1+ \langle \sigma^z_{} \rangle -
2 	\abs{\langle \sigma^+_{} \rangle}^2 \right).\label{eq:pk}
\end{align}
In \cref{fig:sm_populationa} we plot these steady-state populations as
a function of pump strength, across the transition.  When rescaled by $1/N$, the
\(k=0\) has vanishing population in the normal state and becomes non-zero when
macroscopic coherence arises in the lasing state.

\subsection{Dark exciton spectral weight}

An established signature of excitonic dark states in coupled light-matter
systems is a residual peak in the absorption spectrum at the exciton
energy~\cite{houdr1996,Eastham2001,Herrera2016a,Herrera2016b,Cwik2016,Herrera2018}.
This occurs when either static~\cite{houdr1996,Eastham2001} or
dynamic~\cite{Herrera2016a,Herrera2016b,Cwik2016,Herrera2018} disorder can mix
the bright and dark states.  Mathematically, this arises due to the structure of
the imaginary part of the molecular self-energy \(\Sigma^{-+}\), \cref{eq:smp}.
One finds that the weight of any residual peak decreases as the light-matter
coupling \(\Omega\) increases---one may understand this by considering the
imaginary part of \cref{eq:DR} for which \(\abs{\Sigma^{-+}}^2\propto \Omega^4\)
appears in the denominator.  On the other hand, at small values of \(\Omega\)
the residual peak cannot be separated from the upper and lower polariton.  The
values of $\Omega$ shown in the Letter are in fact too small to separate the
residual peak from the upper polariton.  In \cref{fig:sm_populationb} we show
that by further increasing $\Omega$ this residual dark exciton peak may be
clearly observed.

\section{Calculating dynamics with PT-TEMPO}%

In this section, for completeness, we discuss the TEMPO method introduced in
Ref.~\cite{Strathearn2018}, construction of the PT, and its combination with the mean-field dynamics. 
\new{%
We also discuss the types of problems for which this method may be applied.
}

The TEMPO network is built around a discretized Feynman-Vernon influence
functional~\cite{Feynman1963, Makri1995, Makri1995a}, which captures the effect
of the bath, including memory effects.  Approaches based on the influence
functional require summation over intermediate states.  In TEMPO, this summation
is formulated as the contraction of a tensor network.  To derive the influence
functional tensors and construct this tensor network, the coherent evolution of
the system density operator \(\rho\) from time \(t_0\) to \(t_M\), described by
the total Liouvillian  \(\mathcal{L}(t)=-i[H_S(t)+H_E,\cdot]\), is firstly
divided into \(M\) short-time propagations,
\begin{align}
	\rho(t_M) &= 
	T_{\leftarrow}\exp(\int_{t_0}^{t_M}dt\mathcal{L}(t))\rho(t_0) \nonumber\\
	&= T_{\leftarrow} \prod_{m=0}^{M-1}\exp(\int_{t_m}^{t_{m+1}}dt
	\mathcal{L}(t))\rho(t_0), \label{eq:tordered}
\end{align}
where \(t_m=m\delta t\) and \(T_{\leftarrow}\) time-orders these expressions,
placing earlier times to the right.  Next, the system and environment
contributions at each time step are split up using a symmetrized Suzuki-Trotter
expansion:
\begin{align}
	\begin{split}
		&\exp[-i\int_{t_m}^{t_{m+1}} dt 
		\left( \mathcal{L}_S(t) + \mathcal{L}_E \right)]\\
		&=
\exp[-i \int_{t_m+\delta t/2}^{t_{m+1}} dt \mathcal{L}_S(t)]
\exp(-i \mathcal{L}_E \delta t)\\
&\times\exp[-i\int_{t_m}^{t_m+\delta t/2}dt \mathcal{L}_S(t)]+ 
O(\delta t^3).
	\label{eq:trotter}
\end{split}
\end{align}
Note that \(\mathcal{L}_E=-i[H_E,\cdot]\) is time-independent, but
\(\mathcal{L}_S(t)=-i[H_S(t),\cdot]\) depends on time in general.  In our
problem, \(\mathcal{L}_S\) depends implicitly on time via the time-dependent
expectation value $\langle a(t) \rangle$, \(\mathcal{L}_S=\mathcal{L}_S(\langle
a^{\vphantom{\dagger}}_{}(t)\rangle )\).

\begin{figure}[h]
	\centering
	\includegraphics[scale=.65]{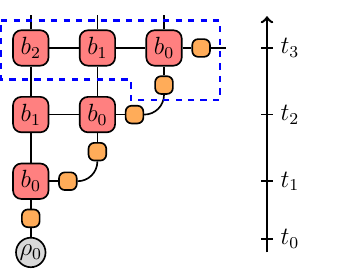}
	\caption{%
		Growth of the tensor network in the TEMPO method according to
		\cref{eq:rhojn}.  \(M\) bath tensors (red) and two system propagators
		(orange) are added at the \(M^{\text{th}}\) time step (here \(M=3\) in
		the blue dashed box).  In practice, a finite memory approximation is
		made in which at most \(K\) bath tensors are added in one step.  The
		initial state---a vector with \(d^2\) elements---is shown as a gray
		circle. 
	}
	\label{fig:tempo}
\end{figure}

The full time evolution is then written as a sum over system states by inserting
a resolution of identity between successive short time propagators in
\cref{eq:tordered}.  To express this sum in tensor notation, it is convenient to
vectorize system operators and matrices using a single index \(j=1,\ldots,
d^2\), where \(d\) is the system Hilbert space dimension (\(d=2\) for the system
we consider), and choose a basis for which the system-environment coupling is
diagonal (\(\sigma^z\) in our model).  In this basis the components of the
system density matrix at \(t_M\) take the form~\cite{Strathearn2018}
\begin{align}
	\begin{gathered}
	\rho_{j_M}(t_M) = \sum_{j_0, j_1,\ldots j_M}\left[ 
		\prod_{m=1}^{M-1} 
		K'_m(j_{m+1}, j_{m})\right. \\
		\left. \times\left( \prod_{k=0}^{m-1} I_k(j_m, j_{m-k}) \right) 
		K_m(j_m, j_{m-1})
	\right] \rho_{j_0}(0)\label{eq:rhojn}
\end{gathered}
\end{align}
where \(K_m(j,j')= \left[T_{\leftarrow} \exp(\int_{t_m}^{t_m+\delta t/2} dt
	\mathcal{L}_S(t))\right]_{j,j'}\) is a two-index object i.e.\ a tensor such
	that contracting \(K_m(j,j')\) with \(\rho_{j'}\) enacts system-only
	evolution over the half-time step \([t_m, t_m+\delta t/2]\), and similarly
	\(K'_m(j,j')= \left[T_{\leftarrow} \exp(\int_{t_m+\delta t/2}^{t_{m+1}} dt
	\mathcal{L}_S(t))\right]_{j,j'}\).  The index $m$ on $K_m, K'_m$ indicates
	the fact that these tensors vary with time step, because of the
	time-dependent system Hamiltonian.  The other objects \(I_k(j,j')\) are the
	bath influence functions that, taken together, capture all possible effects
	of the environment on the system.  For these, the index $k$ indicates the
	time difference over which the bath influence is evaluated.  The bath
	influence function does not depend on the label $m$ as the bath is
	time-independent. These influence functions depend both on the spectrum of
	the environment \(J(\nu)\) and the system operator coupling to the
	environment; see Ref.~\cite{Strathearn2018} for complete expressions.  From
	these we define the bath tensors \(b_k\), 
\begin{align}
	[b_k]_{j,j',\ell,\ell'} = I_k(j,j') \delta_{j,l}\delta_{j',l'},
\end{align}
such that at the \(M^{\text{th}}\) time step, \(M\) bath tensors and two system
propagators may be added to the network according to \cref{eq:rhojn}, as seen in
\cref{fig:tempo}. 

\subsection{Process tensor MPO approach}

\begin{figure}[h]
	\centering
	\vspace{-2\baselineskip}%
	\phantomsubfloat{\label{fig:pta}}
	\phantomsubfloat{\label{fig:ptb}}

	\includegraphics[scale=.65]{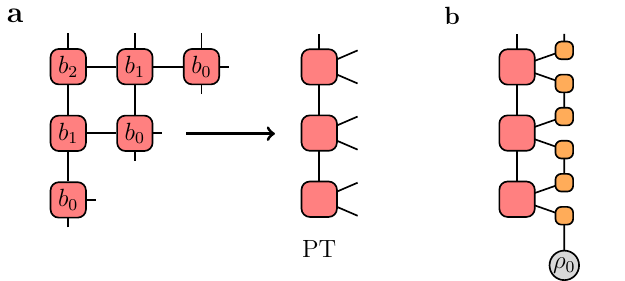}
	\caption{%
		(a) The bath influence tensors of the TEMPO network are contracted
		horizontally to form the process tensor (PT). (b) An initial state
		\(\rho_0\) and system propagators (orange) are then sequentially
		contracted with the PT to calculate the system dynamics. Control
		operations may also be inserted to allow for the measurement of
		multitime correlations.  As in \cref{fig:tempo} time increases in the
		upward vertical direction.
	}
	\label{fig:pt}
\end{figure}

So far we have discussed the tensor network which is common to both the TEMPO
method as originally implemented~\cite{Strathearn2018} and the PT approach.  In
the PT approach, one uses the fact that the bath tensors at each time step may be
contracted independently of the system propagators, as shown in \cref{fig:pta}.
Contracting the tensor network makes use of standard tensor network
techniques~\cite{Orus2014}.   In particular, as the network is contracted,
compression occurs by truncating the singular value decompositions that arise.
After this compression, the storage requirement for the PT is a product of its
length (i.e.\ number of time steps) and the average bond dimension.  We discuss
further below (Convergence of dynamics) how to ensure a small bond dimension and
thus that the PT-MPO to be stored is of a manageable size.

Our PT-MPO runs at a fixed precision \(\epsilon_{\text{rel.}}\)  such that at
each step in the construction of the PT singular values smaller than
\(\epsilon_{\text{rel.}}\) relative to the largest singular value are discarded.
This is in contrast to other MPO methods where instead the  bond dimension of
the tensor is fixed and hence the precision varies.  In addition, a finite
memory approximation is made whereby all correlations are discarded after \(K\)
time steps.

The resulting object, the PT, may be stored and later combined with different
sets of system propagators and initial states to obtain many time evolutions at
relatively little cost.  Additional operators may also be inserted between the
system propagators at this stage for the purpose of calculating multitime
correlation functions~\cite{Pollock2018,Gribben2022:env}.

\subsection{Combining PT-TEMPO with mean-field theory}

In the discussion so far, we described the PT-MPO method for a generic
time-dependent system Hamiltonian $H_S(t)$.  To combine this with mean-field
theory, one then uses a molecular Hamiltonian $H_S(t)=H_i$ from \cref{eq:HMFi}
which depends on $\langle a(t)\rangle$.  To complete the mean-field PT-MPO, one
needs to discretize the evolution of $\langle a \rangle$  consistently with the
system evolution, and thus determine $K_m(j,j^\prime)$, $K_m'(j,j^\prime)$.

Suppose at time \(t_M\) one has a field value  \(\langle
a^{\vphantom{\dagger}}_{} \rangle_M\) and a molecular state \(\rho_M\) with
corresponding spin expectation \(\langle \sigma^- \rangle_M\).  To construct the
system propagators for the next time step, we use the linearization of the field
\begin{align}\label{eq:linearise_a}
	\langle a^{\vphantom{\dagger}}_{} \rangle_M^L(t) = 
\langle a^{\vphantom{\dagger}}_{}\rangle_M	+ (t-t_M)
\left.\partial_t \langle a^{\vphantom{\dagger}}_{}\rangle\right\rvert_{M},
\quad t\in[t_M, t_{M+1}],
\end{align}
where \(\left.\partial_t \langle
a^{\vphantom{\dagger}}_{}\rangle\right\rvert_M\) is the equation of motion
\cref{eq:a} at \(t_M\): 
\begin{align}
	\begin{split}
\left.\partial_t \langle a^{\vphantom{\dagger}}_{}\rangle\right\rvert_M
	&\equiv \left.\partial_t \langle a^{\vphantom{\dagger}}_{}
	\rangle\right\rvert_{\langle 
a^{\vphantom{\dagger}}_{}\rangle_M, \langle \sigma^-\rangle_M}\\
	&= -\left( i\omega_c+\kappa \right)
	\langle a^{\vphantom{\dagger}}_{} \rangle_M - i \frac{\Omega \sqrt{N}}{2}
	\langle \sigma^-\rangle_M \text{.}
\end{split}
\end{align}
Substituting \(\langle a^{\vphantom{\dagger}}_{} \rangle_M^L(t)\) into \(H_S\),
the integrals in \cref{eq:trotter}  may then be performed without further
approximation.

Having applied the total propagator to \(\rho_M\), the remainder of the PT
(describing evolution under \(H_E\) for \(t>t_{M+1}\)) may be traced over to
yield the state \(\rho_{M+1}\) and hence spin expectation \(\langle \sigma^-
\rangle_{M+1}\).  This is used in conjunction with  \(\langle
\sigma^-\rangle_{M}\) to evolve the field from \(t_M\) to \(t_{M+1}\) according
to the second-order prescription
\begin{align}
	\langle a^{\vphantom{\dagger}}_{} \rangle_{M+1} = 
	\langle a^{\vphantom{\dagger}}_{} \rangle_M +
	\frac{1}{2} \delta t \left(k_{M_1}+k_{M_2}  \right)\text{,}
\end{align}
where
\begin{align}
	k_{M_1} = \left.\partial_t \langle a^{\vphantom{\dagger}}_{}\rangle
		\right\rvert_{\langle a^{\vphantom{\dagger}}_{}\rangle_M
		,\langle \sigma^-\rangle_M}\text{,}\\
	k_{M_2} = \left.\partial_t \langle a^{\vphantom{\dagger}}_{}\rangle
		\right\rvert_{\langle a^{\vphantom{\dagger}}_{}\rangle_M
	+ \delta t \cdot k_{M_1}, \langle \sigma^- \rangle_{M+1}}\text{.}
\end{align}

\new{
\subsection{Broader applicability of PT-TEMPO with mean-field theory}

As noted in the Letter, there are two classes of problems for which the
mean-field ansatz is exact as \(N\to\infty\): those with many-to-one coupling
such as considered here, and those with all-to-all coupling~\cite{Mori2013,Carollo2021}. 
Systems with many-to-one coupling most typically arise in the context of cavity QED, including cold atoms in single-mode optical cavities, circuit QED, or molecules in optical cavities as discussed here.  Systems with all-to-all coupling arise in the same contexts, when adiabatic elimination of the cavity mode is possible.  More widely, as discussed below, such all-to-all coupling can become a good approximation in cases when each system couples to many others.

There are many physically relevant situations for which the mean-field theory is not exact but offers a good approximation, and so our method may be applied.  The validity of mean-field approximations has been widely considered in equilibrium condensed matter physics~\cite{chaikin1995}.  In the equilibrium case it is known that for high enough dimensions (i.e. beyond the upper critical dimension, which depends on the problem), mean-field theory can be a good approximation to the problem.  In particular, the effect of fluctuations beyond mean-field theory is controlled by the density of states for low energy modes.  Similar questions have been explored in some open quantum systems. 
These include models of polariton
condensation with multiple modes~\cite{Arnardottir2020} (see \hyperlink{sm:7}{Sec.~7} below for a discussion), non-equilibrium spin models (see e.g.~\cite{lee2013,Jin2016,rota2017,Huybrechts2020} and refs. therein), or cold atoms in multimode cavities~\cite{gopalakrishnan2009emergent,Gopalakrishnan2010}.

As a general principle, the approach described in this work can be applied in any context where one has the following features: (1) One can consider many systems, each of which has its own non-Markovian environment.  (2) These systems couple to each other in a way that can be reasonably approximated by mean-field theory.  i.e., systems couple via collective modes, or couple to many of their neighbours, such that a mean-field approximation may become good.
}

\subsection{Convergence of dynamics}

We next discuss the computational parameters relevant to the process tensor
TEMPO algorithm (hereafter `PT-TEMPO') and provide the values of these
parameters used in the Letter, justified by convergence tests of the dynamics.

\begin{figure}
	\centering
	\vspace{-2\baselineskip}%
	\phantomsubfloat{\label{fig:sm1a}}
	\phantomsubfloat{\label{fig:sm1b}}
	
	\includegraphics[scale=0.395]{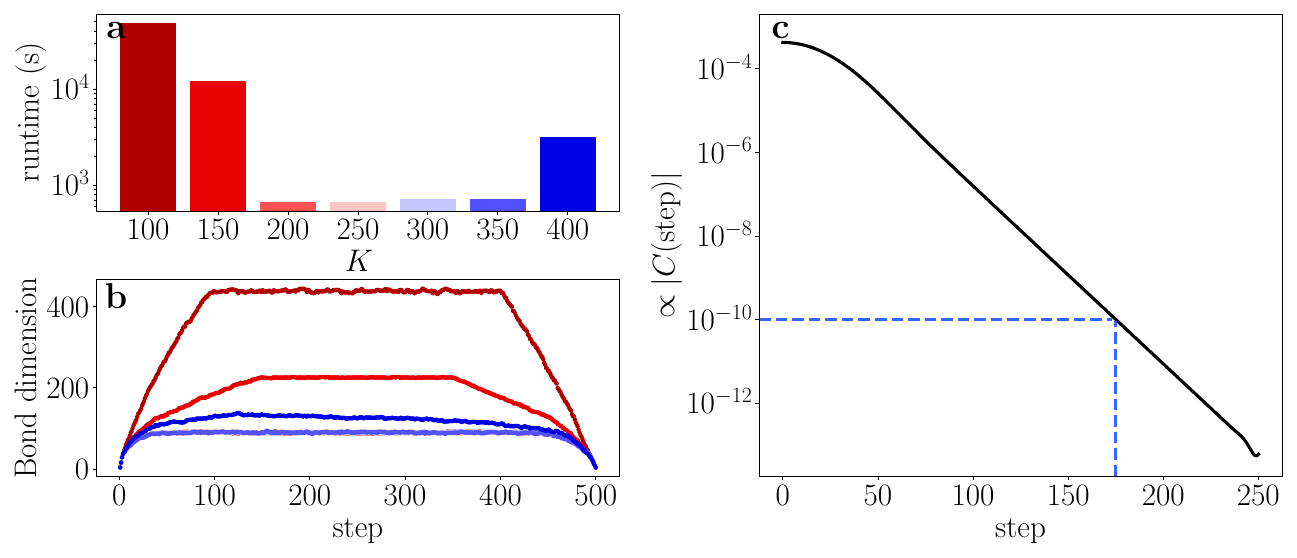}
	\caption{%
		To illustrate the effect of memory cutoff \(K\) on PT computation in the
		PT-TEMPO method we show (a) the total computation time and (b) bond
		dimensions of a PT \(500\) time steps in length for values of  \(K \in
		[100,400]\) (in steps) and a precision of
		\(\epsilon_\text{rel.}=10^{-10}\).  The time step size \(\delta
		t=0.4\)~fs and spectral density parameters matched those of the PT used
		in the Letter at \(T=300\)~K.  Below \(K=200\) a sharp rise in
		computation time is observed corresponding to a growing bond dimension
		across the tensor. These effects grew with decreasing \(K\) such that we
		were unable to construct a PT at \(K<100\) with available resources
		(\(\sim7\)~GB memory).  (c) A hard cutoff  on correlations after \(K\)
		steps corresponds to a discontinuity in the bath autocorrelation
		function \(C(\text{step})\) at \(K\), so we can use the absolute value
		of this function---here scaled such that
		\(\epsilon_{\text{rel.}}=10^{-10}\) coincides with the observed jump in
		computation time at around \(K = 175\)---to estimate the minimum \(K\)
		required to avoid this issue at higher precisions.  This suggests, for
		example, \(K\sim 220\) should be sufficient for the precision
		\(\epsilon_\text{rel.}=5\!\times\!10^{-12}\) used in the Letter. 
	}
	\label{fig:sm1}
\end{figure}

\begin{figure}
	\centering
	\vspace{-2\baselineskip}%
	\phantomsubfloat{\label{fig:sm2a}}
    \phantomsubfloat{\label{fig:sm2b}}
    \phantomsubfloat{\label{fig:sm2c}}
    \phantomsubfloat{\label{fig:sm2d}}
    
	\includegraphics[scale=0.3825]{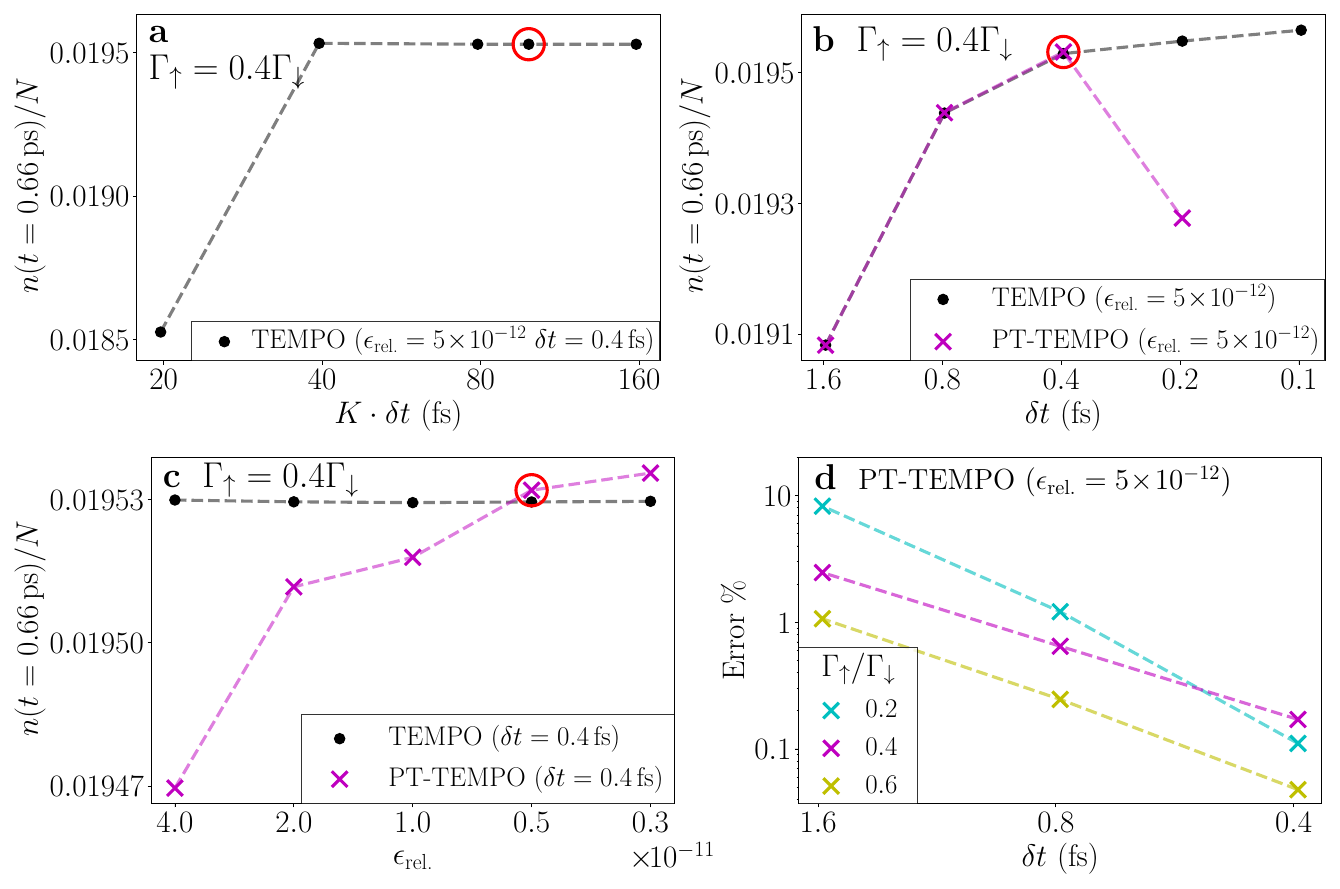}
	\caption{%
		Convergence tests for the computational parameters (a) \(K\), (b)
		\(\delta t\) and (c) \(\epsilon_{\text{rel.}}\).  These panels show the
		\(t=0.66\)~ps value of the scaled photon number \(n/N\) in simulations
		using the PT-TEMPO (crosses) and non-PT TEMPO (filled circles) methods
		at 	\(\Omega=200\)~meV, \(\Delta=20\)~meV,
		\(\Gamma_\uparrow=0.4\Gamma_\downarrow\), and \(T=300\)~K, with losses
		\(\kappa=\Gamma_\downarrow=10\)~meV as in \cref{fig:2}.  In each
		panel, the horizontal axis is ordered so that convergence occurs on
		moving to the right.  In addition,  a red circle indicates data
		corresponding to the computational parameters used in the Letter
		(\(K=250\), \(\delta t=0.4\)~fs,
		\(\epsilon_\text{rel.}=5\!\times\!10^{-12}\)).  (a) The requirement on
		\(K\cdot \delta t\) to attain a manageable bond dimension (see
		\cref{fig:sm1}) means our chosen memory length \(K\cdot \delta t \sim
		100\)~fs is far beyond that at which any significant change in system
		dynamics is observed.  (b) The PT-TEMPO result becomes unstable below
		\(\delta t = 0.4\)~fs whilst the change in the non-PT result continues
		to decrease linearly with time step halvings. (c) The PT-TEMPO method
		appears to require a higher precision (smaller
		\(\epsilon_{\text{rel.}}\)) for comparable accuracy.  This is
		a trade-off of the gain of computational efficiency: the PT-TEMPO data
		point at \(\epsilon_{\text{rel.}}=5\!\times\!10^{-12}\) here took less
		than \(5\) minutes to obtain compared to \(3.5\) hours using the non-PT
		method.  (d) Error in the PT-TEMPO value at
		\(\epsilon_\text{rel.}=5\!\times\!10^{-12}\) for
		\(\Gamma_\uparrow/\Gamma_\downarrow=0.2\), \(0.4\) and \(0.6\) relative
		to non-PT data with the smallest time step \(\delta t=0.1\)~fs at that
		precision.
}
	\label{fig:sm2}
\end{figure}

There are three computational parameters to consider: time step size \(\delta
t\), singular value cutoff \(\epsilon_{\text{rel.}}\) and memory length \(K\).
Evidently \(K \cdot \delta t\) should be chosen to be greater than physical
correlation times of the system.  In fact, we found that if the effective
discontinuity introduced into the bath autocorrelation by truncating the PT
after \(K\) steps was significant on the scale set by
\(\epsilon_{\text{rel.}}\) then a large bond dimension resulted
[\cref{fig:sm1}].  That is, the cutoff effectively implies \(C_{\text{eff}}(t)
= C(t) \Theta(K \delta t - t)\), and the sharp step function leads to the
existence of many singular values of order \(C(K \delta t)\) in the process
tensor.  When \(C(K \delta t) \gtrsim \epsilon_{\text{rel.}}\), this
significantly increases the bond dimension.  At high precisions avoiding this
issue required \(K \cdot \delta t\gtrsim 80\)~fs in excess of any correlation
times in the system and hence the memory cutoff had no effect on the accuracy of
our calculations [\cref{fig:sm2a}].

\cref{fig:sm2b,fig:sm2c} show, respectively, convergence tests under changes in
\(\delta t\) and \(\epsilon_{\text{rel.}}\) where the value of the photon number
\(n/N\) was recorded (crosses) at \(t=0.66\)~ps for one set of system parameters
(\(\Omega=200\)~meV, \(\Delta=20\)~meV, \(\Gamma_\uparrow=0.4\Gamma_\downarrow\)). 
In these panels the data  corresponding to the
computational parameters that were finally chosen, \(\delta t= 0.4\)~fs and
\(\epsilon_{\text{rel.}}=5\times 10^{-12}\), is indicated with a red circle.
For comparison, we include results (filled circles) obtained using the original
(non-PT) implementation~\cite{Strathearn2018} of the TEMPO method.  Note that
the accuracy of the two algorithms for a given set of computational parameters
is not necessarily the same, because of the different ordering of tensor
contractions in the two approaches.  In particular, we noticed the error in the
PT-TEMPO calculation become unstable below \(\delta t= 0.4\)~fs at
\(\epsilon_{\text{rel.}}=5\times 10^{-12}\) [\cref{fig:sm2b}] whilst the non-PT
results remained stable down to \(\delta t = 0.1\)~fs at this precision.  This
issue could not be resolved by further increases in precision, likely due to
operations required to calculate  singular values reaching the limits of machine
(floating point) precision.  Similarly in \cref{fig:sm2c} at  \(\delta
t = 0.4\)~fs we found no benefit in reducing \(\epsilon_{\text{rel.}}\) below
\(5\times 10^{-12}\), instead observing fluctuations in the PT-TEMPO results
about the non-PT value. The discrepancy between the two
implementations did allow us to quantify the error in the PT-TEMPO calculation
at \(\delta t =0.4\)~fs, \(\epsilon_{\text{rel.}}=5\times 10^{-12}\), taking the
\(\delta t = 0.1\)~fs non-PT result as an exact reference. This was done for
three difference pump strengths at \(\Omega=200\)~meV to produce
\cref{fig:sm2d}. By \(\delta t = 0.4\)~fs, the estimated error is well below
\(0.5\%\) in each case.

\subsection{Computational resources used}
For the spectral density \cref{eq:J} (\(\alpha =0.25\), \(\nu_c=150\)~meV) and
temperature \(T=300\)~K, and using the computational parameters described above,
the PT took approximately 3-4 core hours to construct on a \(2.1\)~GHz
Intel\textsuperscript{\tiny\textregistered{}}
Xeon\textsuperscript{\tiny\textregistered{}} processor.  Calculations of similar
length were required to construct PTs for the  the other three temperatures
\(T=250\)~K, \(T=350\)~K and \(T=400\)~K used in \cref{fig:2d}.  Having
precomputed a PT, subsequent contraction with the chosen initial state, system
propagators and control operators took only minutes to complete (we found 10
minutes typical).

\section{Fitting procedures for Figs. 2 and 3}
In this section we detail the 
procedures used to extract the lasing
threshold \(\Gamma_c\) plotted in \cref{fig:2c,fig:2d}.  We also explain how we
check that the steady-state has been reached before applying the operators that
allow us to calculate the two-time correlators used to determine the spectrum 
in \cref{fig:3}.

In order to obtain the steady-state scaled photon number \(n_s/N\) for each set
of system parameters (\(\Omega\), \(\Delta\), \(\Gamma_\uparrow\)) and
environment temperature \(T\), the dynamics were firstly calculated to a final
time \(t_f=1.3\)~ps using a pre-computed PT for that temperature. An exponential
\(a \exp(-b t)+c\) or constant (\(a=b=0\)) fit was then made to the late time
dynamics \(t\geq 1\)~ps.  If the mean squared error of the fit, scaled by the
magnitude of \(a\) (or \(c\) if \(a=0\)), was less than \(10^{-2}\), the fit was
accepted and \(c\) used as the value for \(n_s/N\) (e.g.\ filled circles in
\cref{fig:2b}).  On the contrary, if the error exceeded this cutoff the fit was
deemed poor and the data not used in the subsequent threshold calculation (open
circles in \cref{fig:2b}).  Note in the case  \(n(t_f)/N\) was less than
\(10^{-12}\) no fit was attempted and instead this final value was taken as the
steady-state value.

Before extracting the threshold from the resulting plots of \(n_s/N\) against
\(\Gamma_\uparrow\) such as those in \cref{fig:2b}, it was ensured that there
were sufficient (\(>5\)) values of \(\Gamma_\uparrow\) with valid fits in the
lasing phase.  A quadratic fit of the form \(\Theta(x-\Gamma_c) \left[ a_1
(x-\Gamma_c) + a_2 (x-\Gamma_c)^2 \right]\) was then made to the steady-state
values  at each light-matter coupling, detuning and temperature, yielding the
threshold \(\Gamma_c\) at those parameters; a single point in \cref{fig:2c} or
\cref{fig:2d}.

To produce \cref{fig:3a,fig:3b,fig:3c,fig:3d} the dynamics were calculated to
\(t_f'=1.6\)~ps using the \(T=300\)~K PT (only \(4/5\)ths of this tensor was used
for \cref{fig:2}). Firstly, to reach to steady-state (\(t_f=1.3\)~ps) and,
secondly, to measure either the \(\langle\sigma^+(t)\sigma^-(t_f)\rangle\) or
\(\langle\sigma^-(t)\sigma^+(t_f)\rangle\) correlator (\(t_f\leq t \leq t_f'\)).
These measurements are performed by inserting a control operation \(\sigma^-\)
(or \(\sigma^+\)) in the tensor network at \(t=t_f\) and subsequently recording
the expectation of \(\sigma^+\) (or \(\sigma^-\)).  To ensure the system had
reached the steady-state by \(t_f=1.3\)~ps, the exponential fitting described
above was made  up to \(t_f\);  then only if the fit was valid \textit{and}
close (within \(1\%\) or  \(10^{-5}\) in absolute value) to the observed value
\(n(t_f)/N\) at this time was the state at \(t_f\) deemed suitable for
determining the two-time correlations.

\section{Weak light-matter coupling theory}

In \cref{fig:2c} we included a weak light-matter coupling prediction for the
phase boundary at \(\Omega=100\)~meV.  Here we provide the supporting
calculation and explain its failure to reproduce the observed boundary. This
mismatch is  a consequence of the conditions for lasing being outwith the weak
light-matter coupling regime. Throughout this section ``weak-coupling''
should be interpreted as meaning weak light-matter coupling.

The weak-coupling limit of the model has been considered in
Ref.~\cite{Kirton2015}.  In that paper the authors worked to second order in the
light-matter coupling to derive a weak-coupling master equation of the form
\begin{align}
	\begin{split}
		\partial_t \rho = -i \left[ H_0, \rho \right] + 
		2\kappa \mathcal{L}[a^{\vphantom{\dagger}}_{}]
		&+ \sum_{i=1}^N \left(
	\Gamma_\uparrow \mathcal{L}[\sigma^+_i] 
	+\Gamma_\downarrow \mathcal{L}[\sigma^-_i] \right.\\
	+\Gamma_A(\Delta)  \mathcal{L}[a^{\vphantom{\dagger}}_{}\sigma^+_i]
	 &\left.+\, \Gamma_E(\Delta)  \mathcal{L}[a^{\dagger}_{}\sigma^-_i]\right)
	 \text{,}
\end{split}
\end{align}
where the free Hamiltonian \(H_0=\Delta
a^{\dagger}_{}a^{\vphantom{\dagger}}_{}\) (\(\Delta=\omega_c-\omega_0\)) and \(
\Gamma_{A,E}\) define rates of absorption and emission processes, given by 
\begin{align}
    \label{eq:GammaAE}
	\Gamma_{A,E}(\Delta) = \frac{\Omega^2}{4N} \int_{-\infty}^\infty dt
	e^{\pm i \Delta t}
	\langle \sigma^-(t) \sigma^+(0) \rangle_0 \text{.}
\end{align}
Here \( \langle \sigma^-(t) \sigma^+(0) \rangle_0\) is the correlator for a free
molecule i.e.\  measured in the absence of light-matter coupling.  In
Ref.~\cite{Kirton2015}, to calculate these quantities, it was assumed that the
vibrational environment relaxes fast.  This means that \cref{eq:GammaAE} can be
calculated starting from an equilibrium state of the molecules, an approximation
known as Kasha's rule~\cite{kasha1950characterization}.  For our parameters,
this approximation does not necessarily hold (except for the special case of
\(\Gamma_\uparrow=0\)), so we use the PT-MPO method applied to an
individual molecule to calculate \(\Gamma_{A,E}\).

By making the mean-field factorization approximation, as discussed above,
one can assume \(\langle a^{\dagger}_{} a^{\vphantom{\dagger}}_{
} \sigma^+\sigma^- \rangle \approx \langle a^{\dagger}_{}
a^{\vphantom{\dagger}}_{}\rangle \langle \sigma^+\sigma^- \rangle\) between the
photon number and spin operators. The resulting equation of motion for \(n=
\langle a^{\dagger}_{}a^{\vphantom{\dagger}}_{}\rangle\) is
\begin{multline}
	\partial_t n = -2 \kappa n + 
	N \left[ \Gamma_E(\Delta) (1+n) \langle \sigma^+\sigma^-\rangle
	\right.\\\left.- \Gamma_A(\Delta) n
(1-\langle \sigma^+\sigma^-\rangle) \right]\text{.}
	\label{eq:n}
\end{multline}
At threshold (\(\Gamma_\uparrow=\Gamma_c\)), the coefficient of \(n\) on the
right-hand side of this equation changes from negative to positive. Combining
this with the steady-state population of excited molecules, \(\langle \sigma^+
\sigma^-\rangle = \Gamma_\uparrow/(\Gamma_\uparrow+\Gamma_\downarrow)\), we have
the critical condition
\begin{align}
	-2 \kappa + 
	N \left[ \Gamma_E(\Delta)  \frac{\Gamma_c}{\Gamma_\downarrow+\Gamma_c} 
	- \Gamma_A (\Delta) 
\frac{\Gamma_\downarrow}{\Gamma_\downarrow+\Gamma_c} \right] &=0 \text{,}
\end{align}
from which
\begin{align}
	\frac{\Gamma_c}{\Gamma_\downarrow} =
	\frac{2\kappa + N \Gamma_A(\Delta)}{N \Gamma_E(\Delta) - 2\kappa}\text{.}
	\label{eq:weak_threshold}
\end{align}

Since the rates \(\Gamma_{A,E}\) themselves depend on \(\Gamma_\uparrow\)
through \( \langle \sigma^-(t) \sigma^+(0) \rangle_0\), we solved
\cref{eq:weak_threshold} iteratively for \(\Gamma_\uparrow=\Gamma_c\), taking
advantage of the efficiency with which many sets of system dynamics can be
computed using a single PT. Setting \(\Omega=100\)~meV, at each step
\(\Gamma_\uparrow\) was incremented and \(\Gamma_{A,E}(\Delta)\) evaluated on
the range \(\Delta \in [-100,-20]\)~meV.  The first time equality resulted
between the two sides of \cref{eq:weak_threshold} for a particular \(\Delta\)
provided \(\Gamma_c(\Delta)\)  and hence a single point on the weak-coupling
phase boundary in \cref{fig:2c}.

As is visible in \cref{fig:2c}, even at the smallest \(\Omega\) used, the
weak-coupling theory does not match the predictions of the full model.  Reducing
\(\Omega\) much further leads to a regime where lasing never occurs---the
collective cooperativity becomes too small~\cite{Meiser2009:Propsects}.  As
such, to verify that the full model does match the weak-coupling predictions, we
must consider a different method of comparison.  We choose to do this by
comparing the photon absorption rates of unexcited molecules.   This can be done
by setting \(\Gamma_\uparrow=0\) and preparing an initial state with unexcited
molecules and a small photon field.  We then compare the rates at which this
field decays.

\begin{figure}
	\centering
	\vspace{-2\baselineskip}%
	\phantomsubfloat{\label{fig:weaka}}
    \phantomsubfloat{\label{fig:weakb}}
    
	\includegraphics[width=\linewidth]{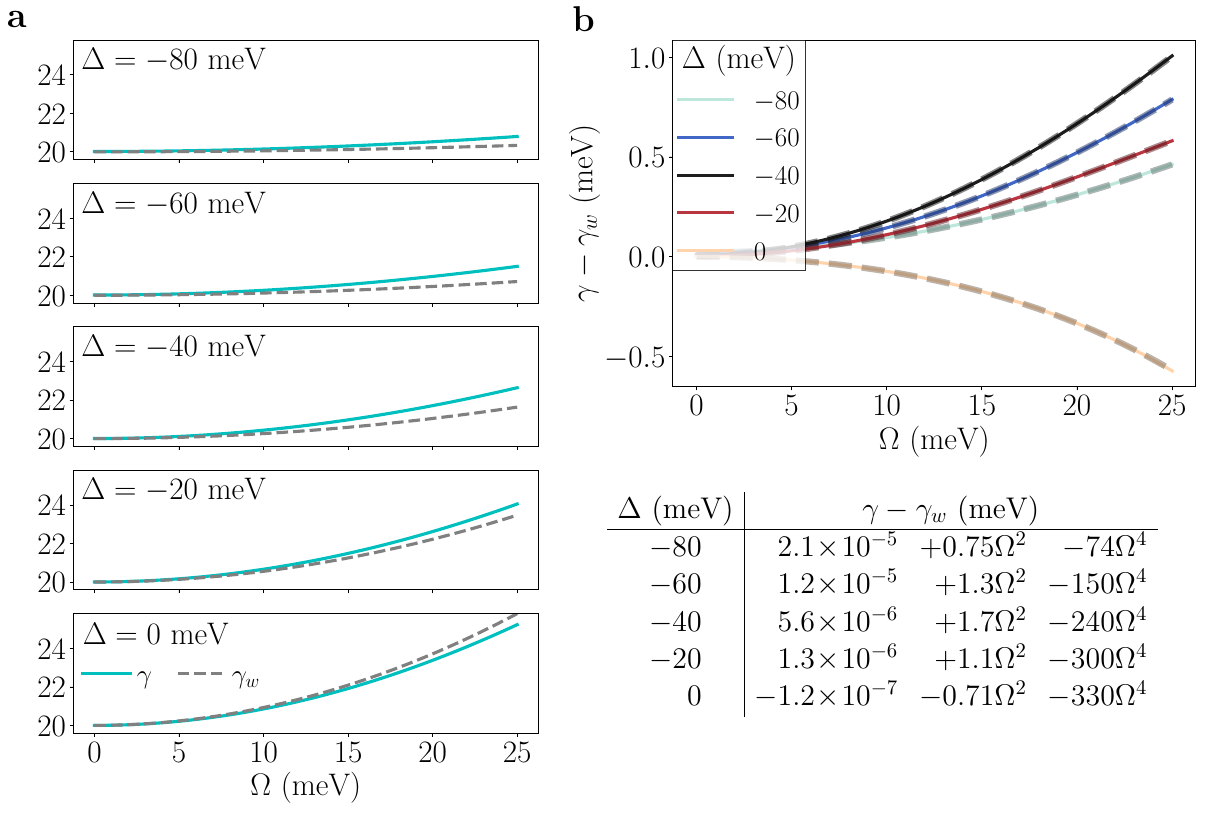}
	\caption{(a) 
		Dependence of effective decay rate \(\gamma\) (cyan) on light-matter
		coupling \(\Omega\) for five different detunings when
		\(\Gamma_\uparrow=0\).  The initial conditions and all other parameters
		were the same as used to produce \cref{fig:2c} (in particular
		\(2\kappa=20\)~meV\(=\gamma(\Omega=20)\)).  The weak-coupling prediction
		\(\gamma_w\) for the rate, \cref{eq:gammaw}, is indicated with a gray
		dashed line.  (b) The difference \(\gamma-\gamma_w\) at each detuning
		with a quartic fit (dashed) recorded in the table shown.  Numerical
		error contributes a small constant and a small \(\Omega^2\) term; it is
		the fourth-order term that describes behavior beyond the weak-coupling
		theory.  Note the dependence on \(\Omega\) is weaker for more negative
		detunings, providing an explanation for the varying error of the
		weak-coupling prediction for the phase boundary in \cref{fig:2c}.
	}
	\label{fig:weak}
\end{figure}

\Cref{eq:n} provides an effective decay rate \(\gamma_w\) for the photon number.
When \(\Gamma_\uparrow=0\) this is simply
\begin{align}
    \label{eq:gammaw}
	\gamma_w = 2\kappa + N\Gamma_A(\Delta) 
\end{align}
and, since an analytical expression for \(\Gamma_A(\Delta)\) is known for
\(\Gamma_\uparrow=0\) (cf. \cref{eq:C}), we can calculate \(\gamma_w\) exactly
for any \(\Omega\) and \(\Delta\), and compare to the rate \(\gamma\) measured
by recording the early time decay (\(t \in [0,400]\)~fs) of \(n/N\) in a PT-MPO
simulation with the same parameters.  This was done for several different
detunings up to \(\Omega=25\)~meV to produce \cref{fig:weaka}. We see the
observed rate (cyan) deviates from the weak-coupling prediction (gray, dashed)
from  \(\Omega=10\)~meV onwards. The breakdown of the weak-coupling
approximation is made clear in \cref{fig:weakb} where we perform a fourth order
polynomial fit to the difference \(\gamma-\gamma_w\): the dominant \(\Omega^4\)
part, which we note increases with \(\Delta\),  cannot be captured by the
second-order weak-coupling theory.

\section{Comparison to effective Holstein--Tavis--Cummings Model}
\begin{figure}
	\centering
	\vspace{-2\baselineskip}%
	\phantomsubfloat{\label{fig:HTCa}}
    \phantomsubfloat{\label{fig:HTCb}}
    \phantomsubfloat{\label{fig:HTCc}}
    
	\includegraphics[width=\linewidth]{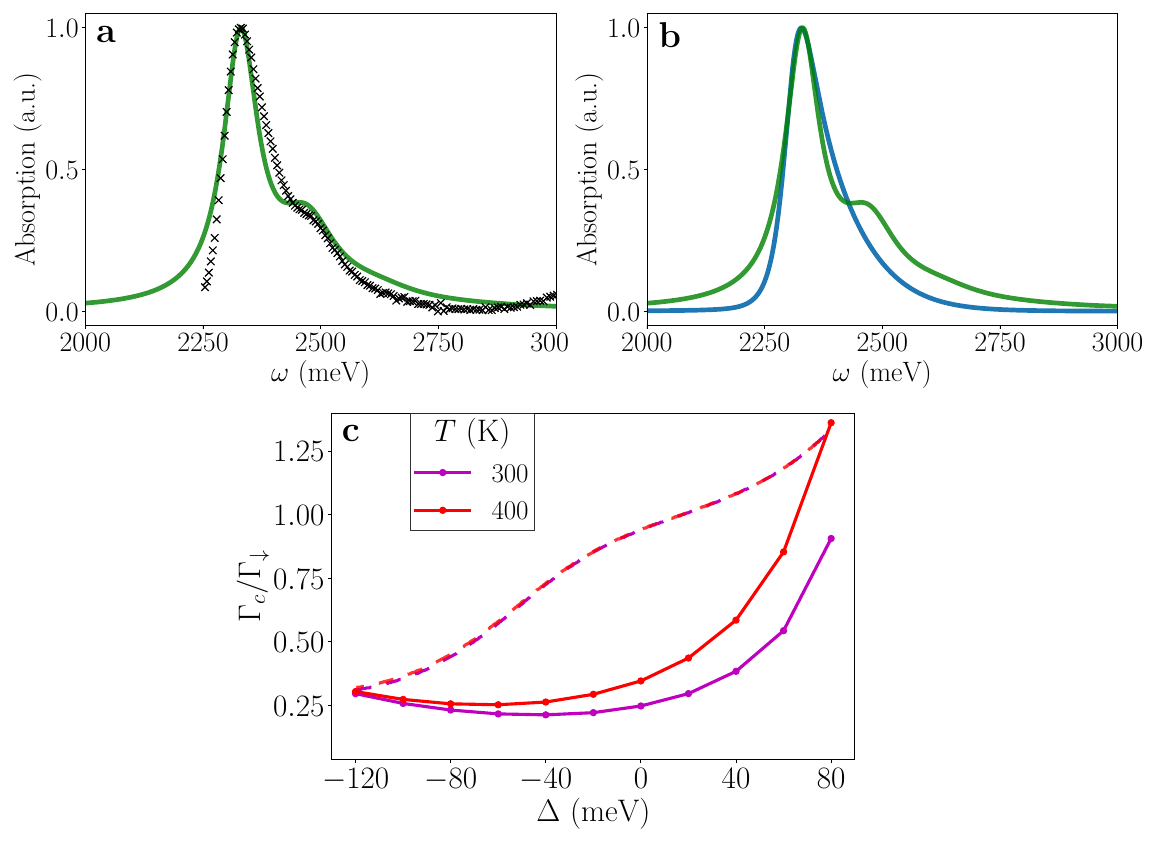}
	\caption{%
		Molecular absorption spectrum of the HTC model, \cref{eq:HTCa,eq:HTCb},
		(green curve) compared to (a) absorption data~\cite{Grant2016} for
		BODIPY-Br at \(T=300\)~K (black crosses) and (b) the spectrum of the
		model considered in the Letter (blue curve).  Note that,  since the HTC
		model has a smaller Stokes shift, a slightly higher two-level system
		frequency \(\omega_0 = 2330\)~meV was	required to match the absorption
		data (\(\omega_0=2310\)~meV for the model in the Letter).  The
		parameters obtained from minimizing the squared deviation of the
		spectrum from the experimental data were \(S=0.12\), \(\Gamma_z=20\)~meV
		and \(\gamma_\nu=60\)~meV.  (c) Lasing threshold
		\(\Gamma_c/\Gamma_\downarrow\) against detuning at \(T=300\)~K and
		\(T=400\)~K.  Dashed lines indicate the phase boundary predicted by the
		HTC model for each temperature, and solid lines those of the model in
		the Letter.  Apart from \(\omega_0\), all other parameters matched those
		used for \cref{fig:2d} (\(\Omega=200\)~meV and
		\(\kappa=\Gamma_\downarrow = 10\)~meV).
	}
	\label{fig:HTC}
\end{figure}
In this section we compare our results to a simplified
model~\cite{Cwik2014,Strashko2018} with a single vibrational mode and find that
the simplified model cannot account for the temperature dependence of the phase
boundary shown in \cref{fig:2d}.

We consider the Holstein--Tavis--Cummings (HTC) Hamiltonian, 
\begin{align}
	\begin{split}
	H_{} &= \omega_c a^{\dagger}_{}a^{\vphantom{\dagger}}_{}
	+ \sum_{i=1}^N \left[ 
	\frac{\omega_0}{2} \sigma^z_{i} + \frac{\Omega}{2\sqrt{N}}
	(a^{\vphantom{\dagger}}_{}\sigma^+_{i}+ a^{\dagger}_{}\sigma^-_{i})
	\right] \\
	&+ \sum_{i=1}^N 
	\omega_\nu \left[ b^{\dagger}_{i} b^{\vphantom{\dagger}}_{i}
		+ \sqrt{S} (b^{\dagger}_{i}+b^{\vphantom{\dagger}}_{i})\sigma^z_i
\right],\end{split}\label{eq:HTCa}
\end{align}
where \(b^{\dagger}_{i}\) creates vibrational excitations of frequency
\(\omega_\nu\) on the \(i^\text{th}\) molecule.  These excitations couple to the
electronic state of the molecule with strength \(\omega_\nu \sqrt{S}\).  Note
that in contrast to Ref.~\cite{Strashko2018} we make the rotating wave
approximation and so do not include a diamagnetic \(A^2\) term. 

Incoherent processes are then included as Markovian terms in
the master equation
\begin{align}
	\begin{split}
	\partial_t \rho &= -i[H,\rho] + 2\kappa \mathcal{L}[a^{\vphantom{\dagger}}_{}]
	+ \sum_{i=1}^N ( \Gamma_\uparrow \mathcal{L}[\sigma^+_i]
	+  \Gamma_\downarrow \mathcal{L}[\sigma^-_i] \\
	&+ \Gamma_z \mathcal{L}[\sigma^z_{i}] 
	+ \gamma_\uparrow \mathcal{L}[b^{\dagger}_{i}+ \sqrt{S} \sigma^z_{i}] 
	+ \gamma_\downarrow \mathcal{L}[b^{\vphantom{\dagger}}_{i}+ \sqrt{S} \sigma^z_{i}]).
	\label{eq:HTCb}
\end{split}
\end{align}
In addition to the pump \(\Gamma_\uparrow\), dissipation \(\Gamma_\downarrow\)
and field decay \(\kappa\) considered in the main text we have introduced dephasing
of the electronic transition at rate \(\Gamma_z\) and vibrational damping. The
latter is due to relaxation of the vibrational mode to thermal equilibrium at
temperature \(T\) with rates \(\gamma_\uparrow = \gamma_\nu n_B(T)\),
\(\gamma_\downarrow = \gamma_\nu (n_B(T)+1)\) where
\(n_B(T)=[\exp(\omega_\nu/T)-1]^{-1}\).  Together these additional processes
approximately describe the effects of the remaining vibrational degrees of
freedom. 

Beyond those parameters that are in common with the model in the Letter there
are then four extra parameters to determine: the vibrational frequency
\(\omega_\nu\), the coupling \(S\), and the rates \(\Gamma_z\) and
\(\gamma_\nu\).  There are several different approaches one might take to decide
these parameters.  We choose to set \(\omega_\nu=140\)~meV according to the
shoulder of the absorption spectrum of BODIPY-Br [\cref{fig:HTCa}]  and proceed
to choose \(S\), \(\Gamma_z\), \(\gamma_\nu\) so as to minimize the sum of
squared deviations of the model's spectrum from the experimental
data~\cite{Grant2016}.  This is consistent with the use of the molecular
absorption data to determine values of the parameters \(\alpha\) and \(\nu_c\)
for the spectral density \cref{eq:J} in the Letter.

In \cref{fig:HTCc} we show the phase boundaries (overlapping dashed lines) for
the HTC model at \(T=300\)~K and \(T=400\)~K, calculated using code publicly
available with Ref.~\cite{Strashko2018}. Alongside we repeat the curves  from
\cref{fig:2d} for the phase boundary of the full model at these temperatures.
While the HTC model does allow for lasing without inversion, the boundary occurs
at a  noticeably higher pump strength over the majority of the region, and has
a minimum controlled  largely by the mode frequency
\(\omega_\nu=140\)~meV~\cite{Strashko2018}.  Most notably, the HTC model shows
no dependence on temperature over the range we consider; this is in marked
contrast to the results of the model described in the Letter.  This occurs
because the relaxation rates \(\gamma_\uparrow\), \(\gamma_\downarrow\) depend
on temperature via the occupation \(n_B=[\exp(\omega_\nu/T)-1]^{-1}\) of the
vibrational mode, but \(\omega_\nu=140\)~meV far exceeds \(T=300\,\text{K}\sim
26\)~meV and  \(T=400\, \text{K}\sim 35\)~meV hence \(n_B(T)\sim0\) for these
and indeed all experimentally relevant temperatures.   In contrast, the approach
described in the main text involves a continuum of low-frequency vibrational
modes; the population of those modes can vary significantly over the relevant
temperature range.  

\section{Multimode model and momentum-dependent spectra}
In this section we discuss the application of our method to an extended model
containing multiple photon modes, and how this allows one to calculate the
$k$-dependent optical spectra shown in \cref{fig:3}.

When including multiple photon modes, the system Hamiltonian becomes
\begin{align}
	\begin{split}
		H_S = &\sum_{\vect{k}} \omega_{c, k} 
		a^{\dagger}_{\vect{k}} a^{\vphantom{\dagger}}_{\vect{k}}
	+\sum_{i=1}^N\biggl[
\frac{\omega_0}{2} \sigma^z_{i}+\biggr.\\
	\frac{\Omega}{2\sqrt{N}} &\sum_{\vect{k}} \biggl.
\left( a^{\vphantom{\dagger}}_{\vect{k}} e^{- i\vect{k} \cdot \vect{r}_i} \sigma^+_{i}
+ a^{\dagger}_{\vect{k}}e^{ i\vect{k} \cdot \vect{r}_i} \sigma^-_{i} 
\right)
	\biggr].	
\end{split}
\end{align}
The form of the mean-field equations in this multi-mode case remains similar to
that presented in the main text.  Indeed, if one assumes that only the $k=0$
photon mode acquires a non-zero occupation, the mean-field equations are
unchanged from those previously considered---the validity of this assumption is
discussed further below.

For the optical spectra, derived from the two-time correlations, we must now
consider momentum-dependent Green's functions $D^{R,K}_{\vect{k}}(\omega)$,
which involve the photon energy $\omega_{c,k}$, and a $\vect{k}$-dependent self
energy.  In a translation-invariant system, this self-energy is diagonal in
momentum and takes the form:
\begin{align}
	\Sigma^{-+}_{\vect{k}}(\omega) 
    = 
    \frac{i\Omega^2}{4N} \sum_{i,j=1}^N
    \int_{0}^\infty dt e^{i\omega t}  \langle
    [\sigma_i^-(t),\sigma_j^+(0)]
    \rangle
    e^{i(\vect{r}_i-\vect{r}_j)
	\cdot\vect{k}}\text{,} \label{eq:smp_k}\\
	    \Sigma^{--}_{\vect{k}}(\omega) 
    = 
    \frac{i\Omega^2}{4N} \sum_{i,j=1}^N
    \int_{-\infty}^\infty dt e^{i\omega t}  \langle
	\{\sigma_i^-(t),\sigma_j^+(0)\}
    \rangle
    e^{i(\vect{r}_i-\vect{r}_j)
	\cdot\vect{k}}\text{.} \label{eq:smm_k}
\end{align}
Below threshold, where the expectations \(\langle \sigma^-_{i}(t)\rangle\),
\(\langle \sigma^+_{j}(0) \rangle\) vanish, only terms with $i=j$ survive within
our mean-field approximation.  We then see the self-energies are independent of
$\vect{k}$ and reduce to those of the single mode model, \cref{eq:smp,eq:smm}.

Above threshold, it is still true that the commutator in \cref{eq:smp_k}
vanishes for \(i\neq j\) within mean-field theory, giving a $k$-independent
expression.  For the anti-commutator in \cref{eq:smm_k} we must now note that
the expectation  \(\langle \sigma^-_{i}(t) \rangle\) is  non-zero. For the
lasing state this term in fact oscillates at the lasing frequency, which we will
denote $\mu$, i.e.\ \(\langle \sigma^-_{i}(t) \rangle = \langle \sigma^-_{i}(0)
\rangle e^{-i \mu t}\).  When lasing occurs at $\vect{k}=0$, this expectation is
identical on all sites, so the anti-commutator expectation takes the form:
\begin{equation}
\label{eq:Aconn}
\langle	\{\sigma_i^-(t),\sigma_j^+(0)\} \rangle
=
2|\langle \sigma^- \rangle|^2e^{-i \mu t} + \mathcal{A}_c(t)\delta_{ij}
\end{equation}
where $\mathcal{A}_c(t)=\langle	\{\sigma_i^-(t),\sigma_i^+(0)\} \rangle
- 2|\langle \sigma^- \rangle|^2$ is the connected part of the expectation. Here
we have used the fact that within  mean-field theory, the connected part exists
part only for $i=j$.  Using \cref{eq:Aconn} in \cref{eq:smm_k} we find:
\begin{multline}
    \Sigma^{--}_{\vect{k}}(\omega) 
    = 
    \frac{i\Omega^2}{4} \biggl[
    2 \pi N 
    \delta_{\vect{k},0} 
    \delta(\omega-\mu)
    2|\langle \sigma^- \rangle|^2
    \biggr.\\\biggl.+
    \int_{-\infty}^\infty dt e^{i\omega t}  
	\mathcal{A}_c(t) \biggr]
	\text{.} \label{eq:smm_k_eval}
\end{multline}
The first term here is the source of the delta-singularity seen in the
photoluminesnce spectrum in \cref{fig:2c}. This singularity exists only at the
lasing wavevector, here taken to be $\vect{k}=0$.

We conclude this section by addressing the validity of a mean-field plus
fluctuation treatment for the multimode model.  As has been discussed
extensively (see e.g. Ref.~\cite{Arnardottir2020,Keeling2020}), such a treatment is
valid provided the number of molecules is large compared to the number of
relevant photon modes---those with energies sufficiently close the molecular
transition energy. 

To make this concrete, consider a finite system of area $A$.  Denoting the areal
density of molecules by \(\rho\),  the number of molecules is \(\rho A\).  To
count photon modes, we use the mode spacing \(k=2\pi/\sqrt{A}\) , and count the
number of modes with energy less than \(E\):
\(N_{\text{ph}}=m_{\text{ph}}AE/(2\pi)\) (recall \(\hbar=1\)).  Hence the number
of molecules per relevant photon mode is \(N/N_{\text{ph}} = E_\rho/E\) where
\(E_\rho =2\pi \rho/m_{\text{ph}}\).  For typical molecular
densities~\cite{eizner2019} we find \(E_\rho\sim 10^7\)~eV. This is many orders of
magnitude greater than any relevant energy scale in the problem, notably the
Rabi frequency \(\Omega \sim 100\)~meV. Therefore there are indeed many more
molecules than relevant photon modes, so the mean-field plus fluctuation
treatment is expected to be accurate.

A separate question for a multimode model is whether it is indeed the
$\vect{k}=0$ mode which condenses.  This question, which is beyond the scope of
this work, is discussed in~\cite{Strashko2018,Arnardottir2020} for the simpler
Holstein-Tavis-Cummings model.  It is found there that for $\Delta>0$,
condensation in $\vect{k}=0$ near threshold is typical.

\section{Inverse Green's functions in the normal state} %
\begin{figure}
	\centering
	\vspace{-2\baselineskip}%
	\phantomsubfloat{\label{fig:additional_pla}}
    \phantomsubfloat{\label{fig:additional_plb}}
    \phantomsubfloat{\label{fig:additional_plc}}
    
	\includegraphics[width=\linewidth]{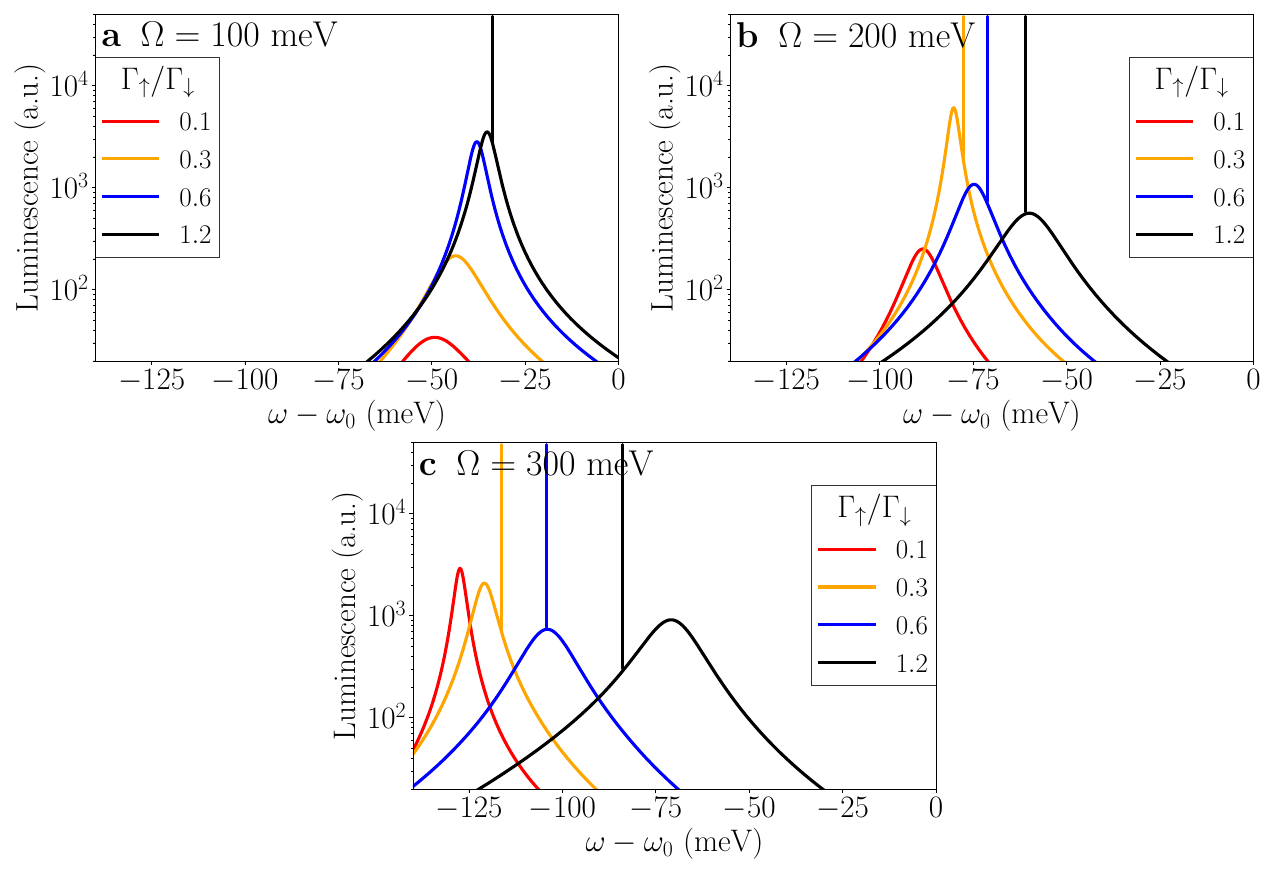}
	\caption{%
		Photoluminescence \cref{eq:pl} at \(k=0\) for four different pump
		strengths when (a) \(\Omega=100\)~meV, (b) \(\Omega=200\)~meV (repeat of
		\cref{fig:3c}) and (c) \(\Omega=300\)~meV. All other parameters match
		those used in \cref{fig:3c}.  A vertical line indicates a lasing peak in
		the spectrum.  Note that, at \(\Omega=100\)~meV, only the highest pump
		strength considered, \(\Gamma_\uparrow=1.2\Gamma_\downarrow\) is
		sufficient to induce lasing. Increasing the light-matter coupling both
		reduces the threshold and redshifts the spectrum.
	}
	\label{fig:additional_pl}
\end{figure}

In this section we examine the inverse retarded and Keldysh Green's functions
below threshold which provide insight into the normal state excitation spectra
and distributions. For reference we show in  \cref{fig:additional_pl} the
photoluminescence \(\mathcal{L}_{\vect{k}=0}(\omega)\), \cref{eq:pl}, at
different pump strengths for light-matter couplings \(\Omega=100\)~meV and
\(\Omega=300\)~meV, in addition to the panel at \(\Omega=200\)~meV included in
the Letter. To simplify the discussion, we work at \(\vect{k}=0\) throughout
this section.

Insight into the normal state excitation spectra and distributions is provided
by studying the components of the inverse Green's functions.  We may define
the components \(A(\omega)\), \(B(\omega)\), \(C(\omega)\) via
\begin{align}
	\left[ D^R(\omega) \right]^{-1} &= A(\omega)+iB(\omega)\text{,}\label{eq:DRIab}\\
	\left[ D^{-1}(\omega) \right]^{K} &= i C(\omega)\text{,}\label{eq:DIKc}
\end{align}
where \(\left[ D^{-1} \right]^{K} \) is such that \( D^K = - D^R \left[ D^{-1}
\right]^{K} D^A\).  The spectral weight (density of states)
\(\rhospec(\omega)=-2 \Im D^R(\omega)\) and mode occupation function
\(2n(\omega)+1= i D^K(\omega)/\rhospec(\omega)\) may then be
written~\cite{Szymanska2007}
\begin{align}
	\rhospec(\omega) &= \frac{2B(\omega)}{A^2(\omega)+B^2(\omega)}\text{,}\\
	n(\omega) &= \frac{1}{2}\left[ \frac{C(\omega)}{2B(\omega)}-1 \right]\text{,}
\end{align}
and the photoluminescence 
\begin{align}
	\mathcal{L}(\omega) = \frac{C(\omega)-2B(\omega)}{2 
	\left[ A(\omega)^2+B(\omega)^2 \right]} \equiv \rhospec(\omega)n(\omega)
	\text{.}
\end{align}
The function \(B(\omega)\) has the role of an effective linewidth for the normal
modes whose position is determined by the zeros of \(A(\omega)\). In the absence
of light-matter coupling (\(\Sigma^{-+}\equiv\Sigma^{--}\equiv0\)), \(B(\omega)
= \kappa\) is a constant and \(A(\omega)=\omega-\omega_c\).  In general it is
possible for the distribution to diverge as \(n(\omega)\sim
1/(\omega-\omega^*)\), where \(\omega^*\): \(B(\omega^*)=0\) defines an
effective boson chemical potential, while the luminescence remains finite.
Instead a condition for a divergence of \(\mathcal{L}(\omega)\), i.e.\
a transition from the normal state to the lasing state, is a simultaneous zero
of \(A(\omega)\) and \(B(\omega)\).

\begin{figure*}
	\centering
	\includegraphics[scale=0.7]{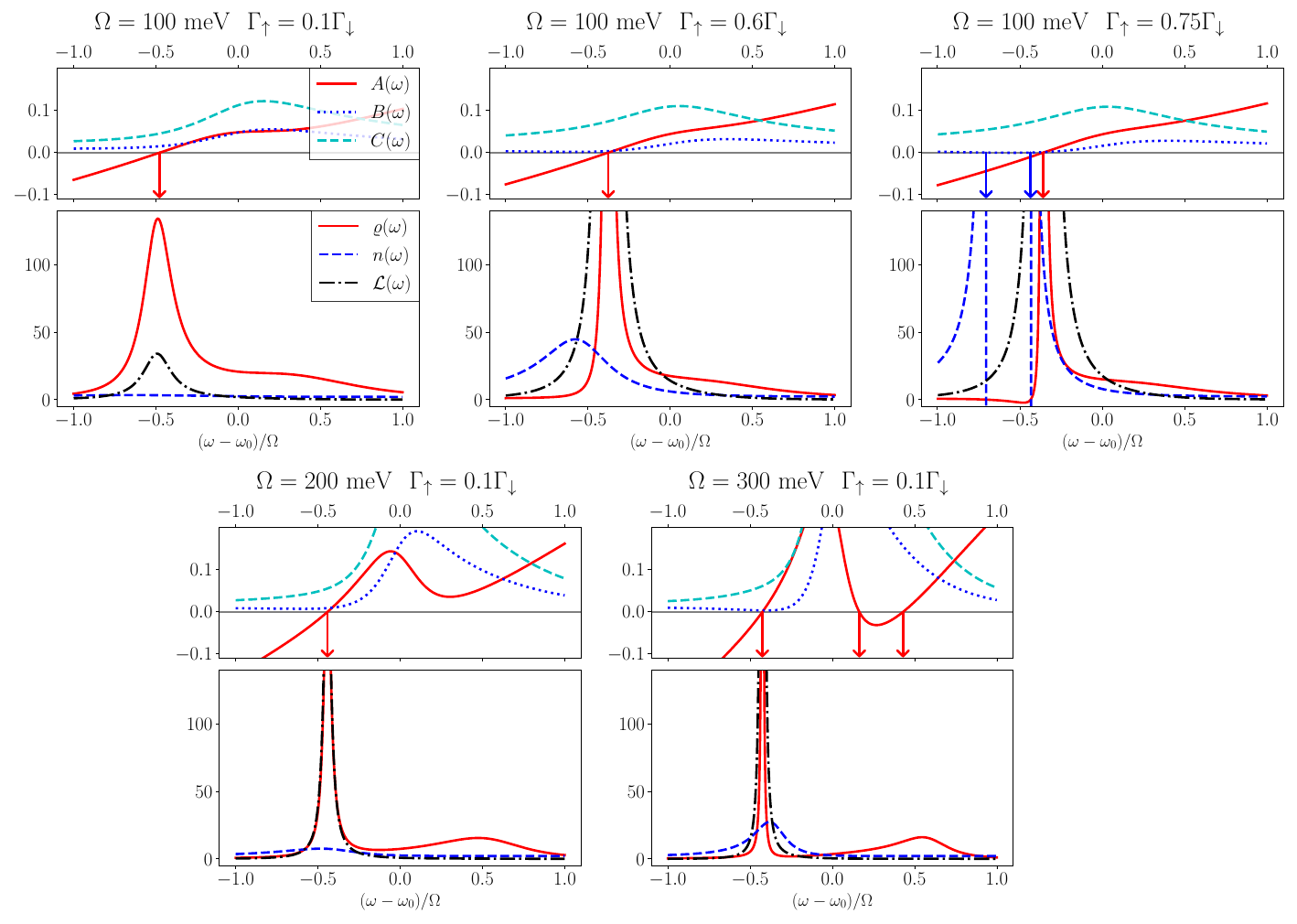}
	\caption{%
		Real and imaginary parts of the inverse retarded and Keldysh Green's
		functions (top axis in each panel) as defined in
		\cref{eq:DRIab,eq:DIKc}  and the corresponding spectral weight,
		occupation and photoluminescence (bottom axis).  Top row:
		\(\Gamma_\uparrow/\Gamma_\downarrow=0.1\), \(0.6\), \(0.75\) at
		\(\Omega=100\)~meV (\(\Delta=-20\)~meV and \(T=300\)~K).  The first two
		pump strengths correspond to the red and blue curves in
		\cref{fig:additional_pla}. The third,
		\(\Gamma_\uparrow=0.75\Gamma_\downarrow\), consists of separate data
		obtained using the non-PT TEMPO method (a longer time \(t_f\sim16\)~ps was
		required to reach the steady-state at this \(\Gamma_\uparrow\) and it
		was more efficient to perform a one-off calculation than compute an
		additional, longer PT).  Red and blue arrows indicate, respectively,
		zeros of the real and imaginary parts \(A(\omega)\) and \(B(\omega)\) of
		\(\left[ D^R \right]^{-1}\).  As the 	threshold
		\(\Gamma_c=0.81\Gamma_\downarrow\) (see \cref{fig:2c}) is approached the
		imaginary part \(B(\omega)\) decreases and develops two zeros (blue
		arrows). Of these, the rightmost is bound to reach the zero of
		\(A(\omega)\) at \(\Gamma_c\), at which point there is a real value
		\(\omega^\ast\) such that \(A(\omega^\ast)=B(\omega^\ast)=0\), signaling
		instability of the normal state~\cite{Szymanska2007,Keeling2010}.
		Bottom row: \(\Gamma_\uparrow/\Gamma_\downarrow=0.1\) at
		\(\Omega=200\)~meV and \(\Omega=300\)~meV.  Note \(A(\omega)\) has two
		additional zeros at \(\Omega=300\)~meV, a feature often taken to signal
		the strong coupling regime. Although the occupation function for this
		light-matter coupling is peaked on the right side of the first zero of
		\(A(\omega)\) here, one expects this will move to the other side before
		the threshold (now at \(\Gamma_c=0.12\Gamma_\downarrow\)) is reached.
	}
	\label{fig:sm3}
\end{figure*}

In the top row of \cref{fig:sm3} we show the components \(A\), \(B\) and \(C\),
as well as the derived \(\rhospec\), \(n\) and \(\mathcal{L}\) as a function of
$\omega$ at \(\Omega=100\)~meV for three pump strengths
\(\Gamma_\uparrow/\Gamma_\downarrow=0.1\), \(0.6\) and \(0.75\) below threshold
at \(\Delta=-20\)~meV (\(\Gamma_c=0.81\Gamma_\downarrow\) from \cref{fig:3c}).
As \(\Gamma_\uparrow\) is increased we see the onset of a divergence in
\(n(\omega)\), which is established \textit{before} the transition, as the graph
of \(B(\omega)\) (blue dotted line) moves downwards to develop two zeros (blue
arrows), one of which is just left of the zero of \(A(\omega)\) (red arrow).

At higher light-matter coupling strengths \(\Omega=200\)~meV and \(300\)~meV
(bottom row of \cref{fig:sm3}), the approach to the transition follows the same
narrative albeit with more spectral weight---including additional zeros of
\(A(\omega)\) at \(\Omega=300\)~meV---at the upper polariton
\(\sim(\omega-\omega_0)/\Omega=0.5\).

\end{document}